\documentclass[12pt]{article} %***
\usepackage{array,epsfig,fancyhdr,rotating}
\usepackage[]{hyperref}  %<----modified by Ivan
\usepackage{amssymb}
\usepackage{diagbox}

\usepackage{array,booktabs}

\usepackage{xcolor}%,dsfont}

\usepackage{amsfonts,enumerate, comment}
\usepackage{apalike}

\usepackage{enumerate}
\usepackage{amsmath}
\usepackage{amsthm}
\usepackage{mathtools}                  % an extension package to amsmath
\mathtoolsset{showonlyrefs}             % the cited refs are shown and the uncited are not shown
\usepackage{dsfont}

\usepackage{tikz}
\usepackage{subcaption, ragged2e,lscape}
\usepackage{multirow}             %for the table

\usepackage{natbib}

\usetikzlibrary{patterns}

\newtheorem{example}{Example}

\def\R{\mathbb{R}}

\def\rb{\textnormal{\textbf{r}}}
\def\ub{\textnormal{\textbf{u}}}
\def\xb{\textnormal{\textbf{x}}}
\def\yb{\textnormal{\textbf{y}}}
\def\zb{\textnormal{\textbf{z}}}
\def\Bb{\textnormal{\textbf{B}}}

\def\Xb{\textnormal{\textbf{X}}}
\def\Wb{\textnormal{\textbf{W}}}

\def\Zb{\textnormal{\textbf{Z}}}
\def\1b{\textnormal{\textbf{1}}}

\def\betab{{\boldsymbol\beta}}

\def\lambdab{\boldsymbol{\lambda}}

\def\var{{\rm var}}
\def\E{{\rm E}}
\def\cov{{\rm cov}}
\def\P{{\rm pr}}

\def\eb{\hbox{$\bf e$}}

\def\xb{\hbox{$\bf x$}}
\def\yb{\hbox{$\bf y$}}

\def\blueR1#1{\textcolor{blue!60}{#1}}
\definecolor{lightgrey}{RGB}{197,200,204}
\definecolor{green}{rgb}{0.55, 0.71, 0.0}

\title{Quasi-Model-Assisted Estimators under Nonresponse \\ in Sample Surveys}

\date{}
\author{Esther Eustache$^*$ and Caren Hasler$^+$\\
\small{$^*$Institut des Sciences du Sport de l'Université de Lausanne, Lausanne, Switzerland} \\
\small{$^+$University of Neuchâtel, Institute of Statistics, Neuchâtel, Switzerland}}

\begin{document}

\renewcommand{\baselinestretch}{2}

%%%%%%%%%%%%%%%%%%%%%%%%%%%%%%%%%%%%%%%%%%%%%%%%%%%%%%%%%%%%%%%%%%%%%%%%%%%%%%%%%%%%%%%%%%%%%%%%%%%%%%%%%%%%%%%%%%%%%%%%%%%%

\fontsize{12}{14pt plus.8pt minus .6pt}\selectfont \vspace{0.7pc}
\centerline{\large\bf QUASI-MODEL ASSISTED ESTIMATORS }
\vspace{1.5pt} 
\centerline{\large\bf UNDER NONRESPONSE IN SAMPLE SURVEYS}
\vspace{.2cm} 
\centerline{Caren Hasler$^{1}$ and Esther Eustache$^2$} 
\centerline{\it $^1$University of Zurich, Department of Psychology, Psychological Methods,}
\vspace{-.4cm} 
\centerline{\it Evaluation and Statistics}
\vspace{-.1cm} 
\centerline{\it $^2$Institute of Sport Sciences of the University of Lausanne, Switzerland} 
 \fontsize{9}{11.5pt plus.8pt minus.6pt}\selectfont

\begin{quotation}
\noindent {\it Abstract:}
  In the presence of auxiliary information, model-assisted estimators rely on a working model linking the variable of interest to the auxiliary variables in order to improve the efficiency of the Horvitz-Thompson estimator. Model-assisted estimators cannot be directly computed with nonresponse since the values of the variable of interest is missing for a part of the sample units. In this article, we present and study a class of quasi-model-assisted estimators that extend model-assisted estimators to settings with non-ignorable nonresponse. These estimators combine a working model and a response model. The former is used to improve the efficiency, the latter to reweight the nonrespondents. A wide range of statistical learning methods can be used to estimate either of these models. We show that several well-known existing estimators are particular cases of quasi-model-assisted estimators. We examine the behavior of these estimators through a simulation study. The results illustrate how these estimators remain competitive in terms of bias and variance when one of the two models is poorly specified.

\par
\end{quotation}

\fontsize{12}{14pt plus.8pt minus .6pt}\selectfont

% ----------------------------------------------------------------
\section{Introduction} \label{section:intro}

In surveys with complete response, the Horvitz-Thompson (HT) estimator is a design-unbiased estimator of population totals \citep{hor:tho:52}. In the presence of auxiliary information, model-assisted estimators can improve the efficiency of the HT estimator by incorporating a working model that links the variable of interest and the auxiliary variables. The roots of model-assisted estimation lie in the Generalized REGression (GREG) estimator introduced by \cite{sar:80} and further examined by \cite{rob:sar:83} and \cite{sar:wri:84}. Over the past decades, there has been a large body of work on model-assisted estimation \citep[e.g.,][]{sar:swe:87,sar:swe:wre:92,fir:ben:98,ful:09:book:sampling}, including recent developments with flexible working models \citep[e.g.,][]{bre:ops:17:modelassist,bre:ops:00,bre:cla:ops:05,bre:ops:joh:ran:07,ops:bre:moi:kau:07,dag:gog:haz:22:model_assisted}.

Nonresponse occurs when the variable of interest is observed for only part of the sample. In this case, neither the HT nor model-assisted estimators can be directly computed, since some values of the variable of interest are missing. A common approach to reducing nonresponse bias is to reweight the respondents in order to compensate for nonrespondents. This produces Nonresponse Weighting Adjusted (NWA) estimators, also called empirical double expansion or propensity-score adjusted estimators \citep[e.g.,][]{sar:swe:87,lun:sar:99,lee:ran:sar:02,bri:13,haz:bea:17:weights:review}. 
Two widely used approaches to estimate the response probabilities are maximum likelihood and calibration.
Maximum likelihood assumes a parametric model for the response probabilities \citep[e.g.,][]{ekh:laa:91,bea:05a,kim:kim:07}, while calibration adjusts weights so that the reweighted estimators of auxiliary variables match known (or design-unbiased estimated) population totals \citep{dev:sar:92,fol:91,dev:dup:93,dup:93,sar:lun:05}. \cite{has:24} introduces a common framework in which the asymptotic behavior of the NWA estimator with response probabilities estimated via maximum likelihood and calibration can be compared. Generalized calibration, also known as instrumental calibration, allows the variables used for response modeling to differ from the variables used for calibration \citep[e.g.,][]{dev:02,kot:06,kot:cha:10,kot:lia:12,les:haz:dha:19}. Nonparametric methods for NWA have also been studied \citep[e.g.,][]{niy:94,niy:97,das:ops:06,das:ops:09}.

In this article, we present and study a class of quasi-model-assisted estimators, which extend model-assisted estimators to the presence of nonresponse by reweighting respondents. Following \cite{bea:05a}, we use the prefix `quasi' to indicate that these estimators are not exactly model-assisted but blend model-assisted and NWA approaches. This class is broad, covering many estimators, including well-known ones.

Quasi-model-assisted estimators rely on two models: the working model links the variable of interest to some auxiliary variables, while the response model links the response probabilities to some auxiliary variables. Each model can be estimated using various statistical learning techniques. For some choices, the resulting estimator can be expressed as a reweighted estimator with weights calibrated to totals of auxiliary variables or to other quantities. Many existing estimators can be viewed as special cases within this class.

Quasi-model-assisted estimators are related to doubly robust (DR) estimators, which use both models and produce consistent estimates if either model is well specified \citep[e.g.,][]{rob:rot:zha:94,rot:rob:sch:98,ban:rob:05,kan:sch:07}. However, the vast majority of pieces of works devoted to DR estimators operate within the standard independent and identically distributed context. Quasi-model-assisted estimators, on the other hand, operate within a survey sampling context. Survey data often come from complex designs, and nonresponse introduces a second phase of implicit sampling with unknown probabilities. Moreover, the target of inference is typically a finite-population parameter rather than a superpopulation parameter. In the context of survey sampling with nonresponse, DR have been studied by \cite{bea:05a} and \cite{kim:haz:14}. Our work extends the class of DR estimators in this context.

A simulation study illustrates that quasi-model-assisted estimators perform well in terms of bias and variance, even when one of the two models (working or response) is misspecified, reflecting their doubly robust property. However, they are not necessarily optimal when both models are misspecified, which is consistent with some observations in \cite{kan:sch:07}.

The article is organized as follows. Section~\ref{section:framework} presents the basic setup, including nonresponse, NWA, and model-assisted estimation. 
Section~\ref{section:quasimodelassisted} introduces quasi-model-assisted estimators, their construction using statistical learning methods, and their property of double robustness. %Section~\ref{section:qmapi} discusses an alternative applicable when auxiliary information is only available at the sample level. 
Section~\ref{section:quasimodelassisted} also contains examples of existing estimators that belong to the class of quasi-model-assisted estimators, illustrating the broad coverage of this class.
Section~\ref{section:simulation} presents simulation results. The article concludes with a discussion in Section~\ref{section:discussion}. %and additional simulation results on real data are provided in the supplementary material.

% ----------------------------------------------------------------
\section{Basic Setup}\label{section:framework}
% ----------------------------------------------------------------

% ----------------------------------------------------------------
\subsection{Survey and Nonresponse}\label{section:nonresponse}
% ----------------------------------------------------------------

We consider a finite population $U=\{ 1,2,...,N \}$ of size $N$. Let $s \subset U$ be a sample of size $n$ selected from $U$ according to a non-informative sampling design $p$. Consider the sample membership indicator $a_k$ of unit $k$ that takes value 1 if unit $k$ is selected in the sample and 0 if not. The first-order inclusion probability of a generic unit $k$ is denoted by $\pi_k$. We have $\P(a_k=1) = \pi_k$, $\P(a_k=0) = 1-\pi_k$, and $\E_p(a_k) = \pi_k$, where subscript $p$ means that the expectation is computed with respect to sampling design $p$. %The covariance between the sample membership indicators is $\Delta_{k\ell} = \mbox{cov}_p(a_k,a_\ell)=\pi_{k\ell}-\pi_k\pi_\ell$.%The first- and second-order inclusion probabilities are denoted, respectively, by $\pi_k$ and $\pi_{k\ell} = \P(k, \ell \in s)$ for generic units $k$ and $\ell$. Consider the sample membership indicator $a_k$ of a unit $k$. We have $\P(a_k=1) = \pi_k$, $\P(a_k=0) = 1-\pi_k$, and $\E_p(a_k) = \pi_k$, where subscript $p$ means that the expectation is computed with respect to sampling design $p(.)$. The covariance between the sample membership indicators is $\Delta_{k\ell} = \mbox{cov}_p(a_k,a_\ell)=\pi_{k\ell}-\pi_k\pi_\ell$.
The goal is to estimate the population total of a variable of interest $y$
\begin{align}
    t = \sum_{k \in U} y_k
\end{align} with values $y_k$ known only for those units in the sample. With no additional information, the total $t$ can be estimated by the HT estimator
\begin{align}
\label{HT}
\widehat{t}_{\pi} &= \sum_{k \in s} \dfrac{y_k}{\pi_k},
\end{align}
which is design-unbiased, i.e., $\E_p(\widehat{t}_{\pi})=t$, if $\pi_k > 0$ for all $k \in U$. %If, additionally, $\pi_{k \ell} > 0$ for all $k,\ell \in U$, a design-unbiased estimator of the variance of $\widehat{t}_{\pi}$ is
%\begin{align}\label{estimator:var:HT}
%  \widehat{\var}\left(\widehat{t}_{\pi}\right) &=  \sum_{k \in s}  \sum_{\ell \in s} \frac{\Delta_{k\ell}}{\pi_{k\ell}} \frac{y_k}{\pi_k}\frac{y_\ell}{\pi_\ell}.
%\end{align}

%In practice, some values $\{ y_k \}$ may be missing because they are collected incorrectly or some units refrain from responding. 
In this work, we suppose that some values $\{y_k\}$ are missing because of non-ignorable nonresponse. Let $p_k$ and $r_k$ denote, respectively, the response probability and the response indicator to variable $y$ of a unit $k \in U$, with $\P(r_k = 1 \mid s) = p_k$ and $\P(r_k = 0 \mid s) = 1-p_k$. Consider $s_r = \{ k \in U \mid a_k = 1, r_k = 1\}$, the set of $n_r$ units in~$s$ for which variable $y$ is known. We refer to the units in $s_r$ as \textit{respondents} and the process generating the respondents as the \textit{response mechanism}.
%The membership indicator of a unit $k \in U$ in the set of respondents $s_r$ is $a_k^* = a_k r_k$. The membership indicator of two different units $k, \ell \in U, k \neq \ell $ in $s_r$ is $a^*_{k\ell} = a_k a_\ell r_k r_\ell$.
%\re{In this paper, we suppose that the response mechanism is independent from the selected sample. XXX Do we really need this assumption, and if yes, conditional or unconditional? See also page 10 and see minor comment 5 of reviewer 2 XXX} %This assumption implies $\E_{p^*} (a_k^*) = \E_p \E_q (a_k r_k) = \pi_k p_k$ and $\E_{p^*} ( a^*_{k \ell}) = \pi_{k \ell} p_k p_\ell $, where subscript $p^*$ means that the expectation is computed with respect to sampling design $p^*(.)$ that selects the set of respondents $s_r$ directly in population $U$.

Let $\xb_k = (x_{k1}, \dots, x_{kq})^\top $ be a vector of $q$ auxiliary variables known for each population unit $k \in U$, and $\zb_k = (z_{k1}, \dots, z_{kt})^\top $ be a vector of $t$ auxiliary variables known for each sampled unit $k\in S$. Auxiliary variables serve various survey purposes, and we differentiate them here based on their role. Values $\{\xb_k, k \in U\}$ are used as predictors of the variable of interest $y$ to improve the efficiency of the HT estimator via model-assisted estimation (see Section~\ref{section:modelassisted}). 
If the auxiliary variables $\{\xb_k\}$ are only known at the sample level, an alternative is presented in Section~\ref{section:qmapi}. 
Values $\{\zb_k, k \in s\}$ are used as predictors of response probabilities to correct for nonresponse bias by reweighting (see Section~\ref{section:NWA}).  These sets of variables may be distinct from one another, partially overlap, or be identical. %Here, we assume they have the same number of variables, i.e., $q=t$. 
We assume that the sampling mechanism is non-informative (or ignorable) and the response mechanism ignorable with respect to the variable of interest conditionally on the auxiliary variables.

% ----------------------------------------------------------------
\subsection{NWA Estimators}\label{section:NWA}
% ----------------------------------------------------------------

Nonresponse can be viewed as a second phase of the survey, as observed by \cite{han:hur:46}. This phase involves selecting a subset of survey respondents from the sampled units according to a response process. In the first phase, a sample~$s$ is selected from population $U$ according to a sampling design $p$. 
In the second phase, a sample $s_r$ is selected from $s$ according to a Poisson sampling design with unknown inclusion probabilities~$\{p_k\}$, called the \textit{response probabilities} \citep{sar:swe:87}. This results in a partition of the sample $s$ into two subsamples: the set of \textit{respondents} and the set of \textit{nonrespondents}. 

If nonresponse is seen as a second phase of the survey, the design weights are multiplied by the inverse of the response probabilities to obtain the \textit{two-phase estimator} or \textit{double expansion estimator}
\begin{equation}
	\widehat{t}_{2\pi} = \sum_{k \in s_r} \dfrac{y_k}{\pi_k p_k}.
\end{equation}
The response probabilities are usually unknown in practice. One possible approach is to postulate a model, called response model, that links $p_k$ to $\zb_k$. Estimated response probabilities $\widehat{p}_k$ are obtained from the fitted response model.
%\re{For instance, we may postulate a parametric model $p_k=1/F(\zb_k^\top\lambdab_0)$ where $F$ is a known function with unknown parameter $\lambdab_0$. Then, we obtain estimated response probabilities $\widehat{p}_k$ from this model, with $\widehat{p}_k=1/F(\zb_k^\top\widehat{\lambdab})$ for some estimator $\widehat{\lambdab}$ of $\lambdab_0$. XXX (on en parle + tard, est-ce qu'on donne tout de même un exemple ?) XXX} 
Finally, the estimated response probabilities replace the true response probabilities in the two-phase estimator. This results in the \textit{NWA estimator}, sometimes also called \textit{empirical double expansion estimator} or \textit{propensity score adjusted estimator},
\begin{equation}\label{estimatornwa}
	\widehat{t}_{NWA} = \sum_{k \in s_r} \dfrac{y_k}{\pi_k \widehat{p}_k}.
\end{equation}

% ----------------------------------------------------------------
\subsection{Model-Assisted Estimators}\label{section:modelassisted}
% ----------------------------------------------------------------

The idea of model-assisted estimation is to postulate a model, sometimes called the \textit{working model} or the \textit{superpopulation model}, that links $y_k$ to $\xb_k$. This model is used to improve the efficiency of the HT estimator while maintaining, or almost, its design unbiasedness. 
We assume that every value $y_k$ in finite population $U$ is a realization of an infinite working model $\xi$, in which
\begin{equation}
    \label{model}
	\xi:~~y_k = m(\xb_k)+\varepsilon_k, ~ k\in U,
\end{equation}
where $m$ is an unknown function, $\E_\xi (\varepsilon_k|\xb_k) = 0$, $\var_\xi (\varepsilon_k|\xb_k)=\sigma_k^2$, and subscript $\xi$ indicates that the expectation and variance are computed under model $\xi$. %Here $\E_\xi (\cdot)$ is a shorcut for $\E_\xi (\cdot|\xb_k)$ and same for the variance. 
Under model $\xi$, the conditional infinite population mean of $y_k$ is $\E_\xi (y_k|\xb_k) = m(\xb_k)$.

The \emph{model-assisted difference estimator} of $t$ is
\begin{equation}
    \label{estimator:diff:m}
    \widehat{t}_{diff} = \sum_{k \in U} m(\xb_k) + \sum_{k \in s} \dfrac{y_k-m(\xb_k)}{\pi_k}.
\end{equation}
%\blueR1{If the population values $\{(\xb_k, y_k), k \in U\}$ were known, the unknown function $m(.)$ could be estimated using a standard statistical method. This would result in a function $m_U(.)$, that is independent of the sample.} 
If only the sample values $\{y_k, k \in s\}$ are known, the unknown function $m$ can be estimated by $\widehat{m}_s$ based on values $\{ (\xb_k, y_k), k \in s \}$. 
Substituting this into~\eqref{estimator:diff:m} yields the \textit{model-assisted estimator}
\begin{equation}
\label{estimator:diff:ms}
\widehat{t}_{ma} = \sum_{k \in U} \widehat{m}_s(\xb_k) + \sum_{k \in s} \dfrac{y_k-\widehat{m}_s(\xb_k)}{\pi_k}.
\end{equation}
\cite{bre:ops:17:modelassist} report that, under some regularity conditions and for some specific working models including heteroscedastic multiple regression, linear mixed models, and some statistical learning techniques, the model-assisted estimator $\widehat{t}_{ma}$ is 1) asymptotically design unbiased and 2) asymptotically more efficient than the HT estimator if the working model provides good approximations $\widehat{m}_s(\xb_k)$ for $y_k$.

% ----------------------------------------------------------------
\section{Quasi-Model-Assisted Estimators}\label{section:quasimodelassisted}
% ----------------------------------------------------------------

% ----------------------------------------------------------------
\subsection{General Class of Estimators}
% ----------------------------------------------------------------

In this paper, we present and study a class of estimators referred to as \textit{quasi-model-assisted estimators}. We adopt the terminology `quasi-model-assisted' from \cite{bea:05a} to emphasize that these estimators do not fully align with the standard definition of model-assisted estimators. 
This class is characterized by the fact that its estimators combine a model-assisted estimator with a NWA estimator. They are constructed as follows.

First, the estimated function $\widehat{m}_s(\cdot)$ is not available in the presence of nonresponse. We use instead an estimator $\widehat{m}_r(\cdot)$ constructed from the respondents values $\{ (\xb_k, y_k), k \in s_r \}$ in the model-assisted estimator \eqref{estimator:diff:ms}. Second, we see nonresponse as a second phase of the survey and reweight the respondents with the inverse of the response probabilities. This gives the \textit{two-phase model-assisted estimator}
\begin{equation}
\label{estimator:diff:mr:U:p}
	\widehat{t}_{2ma} = \sum_{k \in U} \widehat{m}_r(\xb_k) + \sum_{k \in s_r} \dfrac{y_k - \widehat{m}_r(\xb_k)}{\pi_k p_k}.
\end{equation}
This estimator contains the unknown response probabilities $ p_k $. We borrow the idea of the NWA estimation and use instead estimated response probabilities $\widehat{p}_k $ to obtain a \emph{quasi-model-assisted estimator}
\begin{equation}
\label{estimator:diff:mr:U:phat}
	\widehat{t}_{qma} = \sum_{k \in U} \widehat{m}_r(\xb_k) + \sum_{k \in s_r} \dfrac{y_k - \widehat{m}_r(\xb_k)}{\pi_k \widehat{p}_k}.
\end{equation}
The quasi-model-assisted estimator corresponds to a model-assisted estimator where the weights are adjusted for nonresponse. %The applied methodology to obtain the NWA estimator can be seen as a blend between reweighting and imputation as it relies on both a model for the response probabilities and a model that links the variable of interest and the auxiliary variables.
%It covers a wide range of estimators depending on the chosen working model and response model. %We detail some examples in Section~\ref{section:examples}.
A wide class of estimators can be written as \eqref{estimator:diff:mr:U:phat}, including some well-known estimators. Section~\ref{section:existing:DR} presents some examples. In some cases, alternative weights to the design weights $1/\pi_k$ can be used in the right-hand-side of Equation~\ref{estimator:diff:mr:U:phat}. We consider the usual design weights unless otherwise specified. The quasi-model-assisted estimator relies on two models: the \textit{working model} and the \textit{response model}. Section~\ref{section:statistical:learning} and Section~\ref{section:resposne:model} present quasi-model-assisted estimators obtained using various statistical learning methods to estimate the working and response models, respectively.
 
To compute the first term of the quasi–model-assisted estimator in Equation~\eqref{estimator:diff:mr:U:phat}, one typically needs the auxiliary variables $\xb_k$ for all population units $k \in U$.
In Section~\ref{section:qmapi}, we describe an alternative method that applies when the auxiliary variables $\xb_k$ are available only for the sample units $k \in s$.

A postulated model, such as a superpopulation model or a response model, can be misspecified in several ways. Model missecification includes  misspecification of the link function (for instance, we apply linear regression to estimate a non-linear regression function) of a misspecification of the auxiliary variables (for instance, we omit variables that are important). Whatever the type of model misspecification, it can result in poor estimators of the total.
The strength of the class of estimators that can be written as $\widehat{t}_{qma}$ in Equation~\eqref{estimator:diff:mr:U:phat} lies in its double robustness. In fact, it is only necessary for one of the models (the working or response model) to be correctly specified for the estimator to have negligible bias, even if the other model is misspecified (see Appendix~\ref{Appendix:DR} for more details).

% ----------------------------------------------------------------
\subsection{Estimated Working Model}\label{section:statistical:learning}
% ----------------------------------------------------------------

\subsubsection{Generalized Regression}\label{section:linear:working:model}
\label{subsection:greg}

Consider the working model $\E_\xi(y_k \mid \xb_k) = m(\xb_k;\betab) = \xb_k^\top \betab$, 
where the function $m$ is known and the parameter vector $\betab$ is unknown.
Using respondent data $\{(\xb_k, y_k), k \in s_r\}$, parameter $\betab$ can be estimated by
%\begin{align}\label{equation:beta}
%    \widehat{\mathbf{B}}_r &=  \left( \sum_{k \in s_r} \frac{\xb_k \xb_k^\top}{c_k}\right)^{-1} \sum_{k \in s_r} \frac{\xb_k y_k}{c_k},
%\end{align}
%where $c_k$ is any of $1, \sigma_k^2, \pi_k, \pi_k \widehat{p}_k,  \pi_k \widehat{p}_k \sigma_k^2$.
\begin{align}\label{equation:beta}
    \widehat{\mathbf{B}}_r &=  \left( \sum_{k \in s_r} \xb_k \xb_k^\top\right)^{-1} \sum_{k \in s_r} \xb_k y_k,
\end{align}
and we have $\widehat{m}_r(\xb_k) = \xb^\top_k \widehat{\mathbf{B}}_r$.
For this choice of working model, the quasi-model-assisted estimator can be written in weighted form $\widehat{t}_{qma} = \sum_{k \in s_r} w_{k,s_r} y_k$, where 
$$
    w_{k,s_r} = \frac{1}{\pi_k \widehat{p}_k} 
            + \left( \sum_{\ell \in U} \xb_\ell^\top - \sum_{\ell \in s_r}{\frac{\xb_\ell^\top}{\pi_\ell \widehat{p_\ell}}}\right)\left( \sum_{\ell \in s_r} \xb_\ell \xb_\ell^\top\right)^{-1} \xb_k.
$$
%{\allowdisplaybreaks
%\begin{align}
%      \widehat{t}_{qma} &= \sum_{k \in U} \xb^\top_k \widehat{\mathbf{B}}_r + \sum_{k \in s_r} \dfrac{y_k - \xb^\top_k \widehat{\mathbf{B}}_r}{\pi_k \widehat{p}_k}\\
            %&= \sum_{k \in s_r} \dfrac{y_k}{\pi_k \widehat{p}_k} + \left( \sum_{k \in U}\xb_k - \sum_{k \in s_r} \dfrac{\xb_k}{\pi_k \widehat{p}_k}  \right)^\top
            %\left( \sum_{k \in s_r} \frac{\xb_k \xb_k^\top}{c_k}\right)^{-1} \sum_{k \in s_r} \frac{\xb_k y_k}{c_k}\\
%            &= \sum_{k \in s_r}\left\{ \frac{1}{\pi_k \widehat{p}_k} 
%            + \left( \sum_{\ell \in U} \xb_\ell^\top - \sum_{\ell \in s_r}{\frac{\xb_\ell^\top}{\pi_\ell \widehat{p_\ell}}}\right)\left( \sum_{\ell \in s_r} \xb_\ell \xb_\ell^\top\right)^{-1} \xb_k \right\}y_k \\
%        \label{equation:qma:calib}\
%            &= \sum_{k \in s_r} w_{k,s_r} y_k.
%\end{align}
%}
The weights $w_{k,s_r}$ are those of the NWA estimator $1/(\pi_k \widehat{p}_k)$ plus a corrective term induced by the working model. 
%When calibration is applied to estimate the response model with calibration variables $\zb_k = \xb_k$, see Section~\ref{section:calibration}, this corrective term cancels out, see Equation~\eqref{estimating:eqn:calib}. In this case, the quasi-model-assisted estimator coincides with the NWA estimator.
%The weights are free from values $\{\xb_k\}$ in $U\backslash s_r$ except through the population totals $\mathbf{t}^X$. 
This formulation shows that only the auxiliary variables for the respondents, $\{\xb_k, k \in s_r\}$, together with their population totals $\mathbf{t}^X = \sum_{k \in U} \xb_k$, are needed to compute this quasi-model-assisted estimator. 
Moreover, the weights $w_{k,s_r}$ do not depend on $y_k$, and can therefore be used for multiple variables of interest, provided that these variables are observed on $s_r$. 
The weights can also be applied to the auxiliary variables, yielding
%Let $\widehat{\mathbf{t}}^X_{qma}$ be the resulting quasi-model-assisted estimator.
%It comes
\begin{align}
    \left(\widehat{\mathbf{t}}_{qma}^X\right)^\top      
    &= \sum_{k \in s_r} w_{k, s_r} \xb_k^\top
    %&= \sum_{k \in s_r} \frac{\xb_k^\top}{\pi_k \widehat{p}_k} 
    %        + \sum_{\ell \in U} \xb_\ell^\top - \sum_{\ell \in s_r}\frac{\xb_\ell^\top}{\pi_\ell \widehat{p_\ell}}\\
    = \sum_{k \in U} \xb_k^\top. 
\end{align}
Thus, the quasi-model-assisted estimator weights are calibrated to the totals of the auxiliary variable, regardless of the response model.

\subsubsection{ \texorpdfstring{$K$}{K}-Nearest Neighbor}

Consider the working model $\E_\xi(y_k|\xb_k) = m(\xb_k)$ with an unknown function $m$. 
Using the $K$-nearest neighbors method, $\widehat{m}_r(\xb_k)$ is computed as the average of the $y$-values of the $K$ respondents whose auxiliary variables $\xb$ are closest to $\xb_k$ in $\R^q$, according to a chosen distance $d(\cdot,\cdot)$ such as the Euclidean distance. 
Let $(1 ; k), \dots, (n_r ; k)$ denote the respondents reordered by increasing distance to $\xb_k$. Then,
\begin{align}\label{equation:regression:function:knn}
    \widehat{m}_r(\xb_k) = \frac{1}{K} \sum_{\ell=1}^K y_{(\ell ; k)}.
\end{align}

We denote by $\rho(\xb_k) = d(\xb_{(K; k)},\xb_k)$ the maximum distance between $\xb_k$ and the corresponding $K$-nearest neighbors. The estimated regression function in Equation~\eqref{equation:regression:function:knn} can be rewritten
$\widehat{m}_r(\xb_k) 
    = K^{-1} \sum_{\ell \in s_r} \mathds{1}_{ \left\{d(\xb_\ell, \xb_k) <  \rho(\xb_k) \right\}} y_\ell$.
With this formulation, the quasi-model-assisted estimator can be written in weighted form
$$
  \widehat{t}_{qma} %&= \sum_{k \in U} \widehat{m}_r(\xb_k) + \sum_{k \in s_r} \dfrac{y_k - \widehat{m}_r(\xb_k)}{\pi_k \widehat{p}_k}\\
            %&=  \sum_{k \in U} \frac{1}{K} \sum_{\ell \in s_r} \alpha^{r}_{k\ell} y_\ell + \sum_{k \in s_r} \dfrac{y_k}{\pi_k \widehat{p}_k} - \sum_{k \in s_r}\dfrac{1}{\pi_k \widehat{p}_k K} \sum_{\ell \in s_r} \alpha^{r}_{k\ell} y_\ell\\
            %&= \sum_{\ell \in s_r} \left\{ \dfrac{1}{\pi_\ell \widehat{p}_\ell} + \frac{1}{K} \left( \sum_{k \in U} \alpha^{r}_{k\ell}  -  \sum_{k \in s_r}\dfrac{\alpha^{r}_{k\ell}}{\pi_k \widehat{p}_k} \right) \right\}y_\ell.
            %&= \sum_{\ell \in s_r} \left\{ \dfrac{1}{\pi_\ell \widehat{p}_\ell} + \frac{1}{K} \sum_{k \in U}\alpha_{k\ell}\left(1    -  \dfrac{a_kr_k}{\pi_k \widehat{p}_k}\alpha_{k\ell} \right) \right\}y_\ell.
            = \sum_{\ell \in s_r} \left\{ \dfrac{1}{\pi_\ell \widehat{p}_\ell} + \frac{1}{K} \left( \sum_{k \in U} \mathds{1}_{ \left\{d(\xb_\ell, \xb_k) \leq  \rho(\xb_k) \right\}} -  \sum_{k \in s_r}\dfrac{\mathds{1}_{ \left\{d(\xb_\ell, \xb_k) \leq  \rho(\xb_k) \right\}}}{\pi_k \widehat{p}_k} \right) \right\}y_\ell.
$$
The weights consist of the NWA weights $1/(\pi_k \widehat{p}_k)$ plus a corrective term induced by the working model. The corrective term rarely cancels out. 
The weights depend on the auxiliary variables through the distance used to form the neighborhoods. They are calibrated to the population size. 
Indeed, when these weights are applied to estimate the total of the constant variable that takes value 1, the resulting estimator takes value $N$. %$\sum_{\ell \in s_r}\mathds{1}_{ \left\{d(\xb_\ell, \xb_k) \leq  \rho(\xb_k) \right\}} = K$ and  
%the resulting estimator is $\widehat{t}_{qma} = N$. 
%\begin{align}
%  \widehat{t}_{qma} %&= \sum_{\ell \in s_r} \left\{ \dfrac{1}{\pi_\ell \widehat{p}_\ell} + \frac{1}{K} \left( \sum_{k \in U} \mathds{1}_{ \left\{d(\xb_\ell, \xb_k) \leq  \rho(\xb_k) \right\}} -  \sum_{k \in s_r}\dfrac{\mathds{1}_{ \left\{d(\xb_\ell, \xb_k) \leq  \rho(\xb_k) \right\}}}{\pi_k \widehat{p}_k} \right) \right\}\mathcolor{red}{\cdot1}\\
%            &= \sum_{\ell \in s_r}  \dfrac{1}{\pi_\ell \widehat{p}_\ell}
%              + \frac{1}{K}\sum_{k \in U}\sum_{\ell \in s_r} \mathds{1}_{ \left\{d(\xb_\ell, \xb_k) \leq  \rho(\xb_k) \right\}}\\
%            &\quad- \frac{1}{K}\sum_{k \in s_r}\dfrac{1}{\pi_k \widehat{p}_k}\sum_{\ell \in s_r}\mathds{1}_{ \left\{d(\xb_\ell, \xb_k) \leq  \rho(\xb_k) \right\}}\\
%         &= N,
%\end{align}
These weights % of the quasi-model-assisted estimator, when the regression function is estimated using the $K$-nearest neighbors, 
are calibrated to the population size regardless of the response model.

\subsubsection{Local Polynomial Regression}\label{section:workingmodel:localpoly}

Local polynomial regression is studied in the context of model-assisted survey estimation in \cite{bre:ops:00}. Consider a working model $\E_\xi(y_k|x_k) = m(x_k)$ with an unknown function $m$ and a scalar $x_k \in \mathbb{R}$. Function $m$ is approximated locally at $x_k$ by $q$-th order polynomial regression. The model is fitted via weighted least squares with weights based on a kernel function centered at $x_k$. \cite{bre:ops:00} propose and study the model-assisted estimator with a survey weighted estimator of $m$ fitted with the sample data $\{(x_k, y_k), k \in s\}$. Adapting their estimator to the context of nonresponse with available respondents data $\{(x_k, y_k), k \in s_r\}$ yields estimated regression function
\begin{align}
      \widehat{m}_r(x_k)  &= \eb_1^\top \left(\Xb_{rk}^\top\Wb_{rk}\Xb_{rk}\right)^{-1}\Xb_{rk}^\top\Wb_{rk}\yb_{rk} = \boldsymbol{\omega}_{rk}^\top \yb_{r},
\end{align}
where $\eb_j$ is a vector with 1 at the $j$-th coordinate and 0 otherwise,
$$    
    \Xb_{rk}  =\left[1 \quad x_j - x_k \quad \cdots \quad  (x_j - x_k)^q\right]_{j \in s_r},
$$
$$
    \Wb_{rk}  = \mbox{diag}\left\{\frac{1}{k_j h}K\left(\frac{x_j-x_k}{h}\right)\right\}_{j\in s_r},
$$
$$
    \yb_{r}  = \left[y_j\right]_{j \in s_r},
$$
$$
    \boldsymbol{\omega}_{rk}^\top = \eb_1^\top \left(\Xb_{rk}^\top\Wb_{rk}\Xb_{rk}\right)^{-1}\Xb_{rk}^\top\Wb_{rk},   
$$
$k_j$ is either $1$ for all $j \in s_r$ or $\pi_j\widehat{p}_j$, $K$ is a continuous kernel function, and $h$ a bandwidth. 
With this estimated regression function, the quasi-model-assisted estimator can be written in weighted form
\begin{equation}
  \widehat{t}_{qma} %&= \sum_{k \in U} \widehat{m}_r(x_k) + \sum_{k \in s_r} \dfrac{y_k - \widehat{m}_r(x_k)}{\pi_k \widehat{p}_k}\\
            %&=  \sum_{k \in U} \boldsymbol{\omega}_{rk}^\top \yb_{r}  + \sum_{k \in s_r} \dfrac{y_k}{\pi_k \widehat{p}_k} - \sum_{k \in s_r}\dfrac{\boldsymbol{\omega}_{rk}^\top \yb_{r}}{\pi_k \widehat{p}_k} \\
            = \sum_{k \in s_r}\left\{ \dfrac{1}{\pi_k \widehat{p}_k} + \sum_{\ell \in U}\left( 1 - \dfrac{a_\ell r_\ell}{\pi_\ell \widehat{p}_\ell}   \right)\boldsymbol{\omega}_{r\ell}^\top \eb_k  \right\}y_k.
\end{equation}
The weights are those of the NWA estimator $1/(\pi_k \widehat{p}_k)$ plus a corrective term induced by the working model. They are free from values $\{y_k\}$ and could therefore be used for several variables of interest provided that they have observed values on $s_r$. In particular, they can be applied to $\{x_k\}$:
\begin{equation}
      \widehat{t}^X_{qma} = \sum_{k \in s_r} \frac{x_k}{\pi_k \widehat{p}_k} + \sum_{\ell \in U}\boldsymbol{\omega}_{r\ell}^\top \sum_{k \in s_r}  \eb_k x_k -  \sum_{\ell \in s_r} \dfrac{\boldsymbol{\omega}_{r\ell}^\top}{\pi_\ell \widehat{p}_\ell} \sum_{k \in s_r}  \eb_k x_k = \sum_{k \in U}x_k,
\end{equation}
where we used
$$
  \boldsymbol{\omega}_{r\ell}^\top \sum_{k \in s_r}  \eb_k x_k = x_\ell.
$$
The weights are calibrated to the totals of the auxiliary variable regardless of the response model.

% ----------------------------------------------------------------
\subsection{Estimated Response Model}\label{section:resposne:model}
% ----------------------------------------------------------------

\subsubsection{Calibration}\label{section:calibration}

When the response probabilities are estimated using calibration, we assume a parametric model $p_k = 1 \slash {F(\mathbf{z}_k^\top \boldsymbol{\lambda}_0)}$, 
where $F$ is a known function with unknown parameter vector $\boldsymbol{\lambda}_0$, and $\mathbf{z}_k$ is a vector of variables observed for all sampled units, as described in Section~\ref{section:framework}. 
With calibration, the estimator $\widehat{\boldsymbol{\lambda}}$ of $\boldsymbol{\lambda}_0$ is obtained as the solution to the estimating equation
\begin{equation}
    \label{estimating:eqn:calib}
   Q(\lambdab) = \sum_{k \in U} \zb_k - \sum_{k \in s_r} \frac{\zb_k}{\pi_k}F(\zb_k^\top\lambdab) = \mathbf{0}.
\end{equation}

Several choices of $F$ are possible, see \cite{dev:sar:92} and \cite{dev:sar:sau:93}. We present only two of these choices in this article. A first choice is the linear function $F(u) = 1+u$. In this case, the response model is $p_k= 1 / (1+\zb_k^\top\lambdab_0)$. This choice is simple and allows for a closed form for the estimated probabilities. It yields $\widehat{p}_k= 1 / (1+\zb_k^\top\widehat{\lambdab})$ where
\begin{align}
    \widehat{\lambdab} &= \left(\sum_{k \in s_r}\frac{\zb_k \zb_k^\top}{\pi_k}\right)^{-1}\left(\sum_{k \in U} \zb_k - \sum_{k \in s_r} \frac{\zb_k}{\pi_k}\right)
\end{align}
One major drawback of this choice is that the solution may yield negative or larger than~1 estimated response probabilities. A second choice is the logistic response model with $F(u) = \left\{1+\exp(u)\right\}/\exp(u)$ and
\begin{align}
    p_k = \frac{\exp(\zb_k^\top\lambdab_0)}{1+\exp(\zb_k^\top\lambdab_0)}.
\end{align}
This function is commonly used because it yields estimated response probabilities between 0 and 1. However, unlike the first choice, there is no closed form for $\widehat{\lambdab}$. The response probabilities can alternatively be estimated via generalized calibration, allowing the variables in the response model to differ from those used for calibration \citep{dev:98,Sautory2003,kot:06}

%In this case, the estimating equation is
%\begin{align}\label{estimating:eqn:gencalib}
%   Q(\lambdab) &= \sum_{k \in U} \tilde{\zb}_k - \sum_{k \in s_r} \frac{\tilde{\zb}_k}{\pi_k}F(\zb_k^\top\lambdab) = \mathbf{0},
%\end{align}
%where $\tilde{\zb}_k$ is a vector of the same dimension as $\zb_k$.
%Unless otherwise specified, we suppose in what follows that the response probabilities $\widehat{p}_k$ are estimated via Equation~\eqref{estimating:eqn:calib}. 
When no information on the calibration variables is available outside the sample, the estimating Equation~\eqref{estimating:eqn:calib} cannot be applied. 
Section~\ref{section:qmapi} presents an alternative approach to obtain estimated response probabilities in this case. 
Another option is to use maximum likelihood estimation.

\subsubsection{Maximum Likelihood Estimation}\label{section:mle}

Another approach consists of estimating the response probabilities via maximum likelihood. We assume that the units respond independently of one another and that the response probabilities are parametrically modeled such that $p_k=1/F(\zb_k^\top\lambdab_0)$ where $F$ is a known function with unknown parameter~$\lambdab_0$. The estimation of $\lambdab_0$ is the solution to the estimating equation
\begin{align}\label{eqn:mle}
  Q (\lambdab) = \frac{\partial }{\partial \lambdab}  
  \sum_{k \in s} c_k\left[r_k \ln F^{-1}(\zb_k^\top\lambdab) +(1- r_k) \ln\left\{1- F^{-1}(\zb_k^\top\lambdab)\right\}\right]= 0,
\end{align}
for some weights $c_k$, see \cite{kim:kim:07}. Common choices for the weights are 1 or~$\pi_k^{-1}$. When $c_k=1$ usual maximum likelihood estimation is applied. For the logistic response model $F(u) = \{1+\exp (u)\}/\exp (u)$, estimating Equation~\eqref{eqn:mle} becomes
\begin{align}\label{eqn:mle:logistic}
  Q (\lambdab) = \sum_{k \in s} c_k\left\{r_k - F^{-1}(\zb_k^\top\lambdab)\right\}\zb_k = 0. 
\end{align} %Double robustness of the quasi-model assisted estimator when maximum likelihood is applied may be obtained using arguments similar to those presented in Section~\ref{section:asymptotics}. This goes beyond the scope of this research.

\subsubsection{Local Polynomial Regression}\label{section:proba:localpoly}

Non-parametric estimation of the response probabilities using kernel regression is presented in~\cite{das:ops:06} and using local polynomial regression in \cite{das:ops:09}. We present the second one since it is more general. With non-parametric estimation, function $F$ is left unspecified. We take
\begin{align}
      \widehat{p}_k  &= \eb_1^\top \left(\Zb_{sk}^\top\widetilde{\Wb}_{sk}\Zb_{sk}\right)^{-1}\Zb_{rk}^\top\widetilde{\Wb}_{sk}\rb_{s} = \boldsymbol{\phi}_{sk}^\top \rb_{s},
\end{align}
where $\eb_j$ is a vector with 1 at the $j$-th coordinate and 0 otherwise,
$$    
    \Zb_{sk}  =\left[1 \quad \zb_j - \zb_k \quad \cdots \quad  (\zb_j - \zb_k)^q\right]_{j \in s},
$$
$$
    \widetilde{\Wb}_{sk}  = \mbox{diag}\left\{\frac{1}{k_j h}K\left(\frac{\zb_j-\zb_k}{h}\right)\right\}_{j\in s}, ~~~
    \rb_{s}  = \left[r_j\right]_{j \in s},
$$
$$
    \boldsymbol{\phi}_{sk}^\top = \eb_1^\top \left(\Zb_{sk}^\top\widetilde{\Wb}_{sk}\Zb_{sk}\right)^{-1}\Zb_{sk}^\top\widetilde{\Wb}_{sk},   
$$
$k_j$ is either $1$ for all $j \in s$ or $\pi_j$, $K$ is a continuous kernel function, and $h$ a bandwidth. Note that in Section~\ref{section:workingmodel:localpoly}, local polynomial regression is fitted on the respondents because values $y_k$ is available only for  the respondents. In the current section, local polynomial regression is applied to the full sample, since $r_k$ is known for all sampled units.

% ----------------------------------------------------------------
\subsection{Auxiliary Information Available at Sample Level}\label{section:qmapi}
% ----------------------------------------------------------------
Some of the aforementioned methods for estimating either the working model or the response model require knowledge of the auxiliary variables for all population units. Table~\ref{table:x:level} summarizes the minimum level of auxiliary information needed to compute the quasi-model estimator in Equation~\eqref{estimator:diff:mr:U:phat} for each of the aforementioned methods. In this section, we present alternative approaches that can be applied when less auxiliary information is available.

\begin{table}[htb!]
\caption{Minimum level of auxiliary information needed in order to compute the quasi-model estimator with three methods to estimate the working model and three methods to estimate the response model.\label{table:x:level}}
\centering
\fontsize{9}{11}\selectfont
\begin{tabular}[t]{ll}
\toprule
\multicolumn{2}{c}{Working Model}\\
Method & Information needed\\
\midrule
Linear   & $\{\xb_k, k\in s_r\}$, and $\sum_{k \in U} \xb_k$\\
$K$-nearest neighbor & $\{\xb_k, k\in U\} $ \\
Local polynomial regression & $\{\xb_k, k\in U\} $\\
\midrule
\multicolumn{2}{c}{Response Model}\\
Method & Information needed\\
\midrule
Calibration & $\{\zb_k, k\in s_r\}$, and $\sum_{k \in U} \zb_k$\\
%Generalized calibration & $\{(\zb_k, \widetilde{\zb}_k), k\in s_r\}$, $\sum_{k \in U} \widetilde{\zb}_k$\\
Maximum likelihood & $\{\zb_k, k \in s\}$\\
Local polynomial regression & $\{\zb_k, k \in s\}$\\
\bottomrule
\end{tabular}
\end{table}

%The values of ${\xb_k}$ must be known for all population units to calculate the first term of the quasi-model-assisted estimator~\eqref{estimator:diff:mr:U:phat}, except in the case of the linear working model, where knowledge of the population totals alone is sufficient.
%$\sum_{k \in U} \xb_k$.
%In order to compute the first term of the quasi-model-assisted estimator in~\eqref{estimator:diff:mr:U:phat}, it is required to know the values $\{\xb_k\}$ for all units of the population for most working models. An exception is the linear working model for which it suffices to know the population total $\sum_{k \in U} \xb_k$ in order to compute the first term in \eqref{estimator:diff:mr:U:phat}. 
%For the other working models presented in Section~\ref{section:statistical:learning}, the values of the auxiliary variables $\{\xb_k, k \in U\}$ for all population units need to be available. Moreover, it is required to know the population total $\sum_{k \in U} \widetilde{\zb}_k$ to obtain estimated response probabilities via~\eqref{estimating:eqn:gencalib}.
%the values $\{\xb_k\}$ are only available for the sampled units and when the population totals of $\zb_k$ is unknown. 

\subsubsection{Working Model}
Most methods for estimating the working model require auxiliary information for population units outside the sample. 
For instance, a linear working model requires the population total $\sum_{k \in U} \mathbf{x}_k$, whereas methods such as $K$-nearest neighbors or local polynomial regression require the full set of population values $\{\mathbf{x}_k, k \in U\}$. 
However, using a slightly modified version of the quasi-model-assisted estimator in Equation~\eqref{estimator:diff:mr:U:phat}, these methods can still be applied when only the sample values $\{\mathbf{x}_k, k \in s\}$ are available. 
Indeed, replacing the population total in the left-hand side of Equation~\eqref{estimator:diff:mr:U:phat} by its HT estimator yields
\begin{equation}
\label{estimator:diff:mr:U:phat:pi}
	\widehat{t}_{qma,\pi} = \sum_{k \in s} \dfrac{\widehat{m}_r(\xb_k)}{\pi_k} + \sum_{k \in s_r} \dfrac{y_k - \widehat{m}_r(\xb_k)}{\pi_k \widehat{p}_k}.
\end{equation}
This estimation can be computed even when no auxiliary information is available on population units outside the sample.

Some other authors propose or study estimators that can be written as in Equation~\eqref{estimator:diff:mr:U:phat:pi} \citep{bea:05a,kim:haz:14}. More details are provided in Section~\ref{section:existing:DR}.
%study an estimator with a similar structure.
%\cite{bea:05a} employs calibrated imputation to address nonresponse by adjusting the estimated values $\{\widehat{m}_r(\xb_k)\}_{k \in s}$ to meet specific balancing constraints.
%\cite{kim:haz:14}  suppose parametric models for both the response model and the working model and estimate both models simultaneously based on a system of estimating equations. Their procedure results in doubly-robust point and variance estimators. Our proposed approach is more general as models can be estimated separately and/or using non-parametric approaches.

\subsubsection{Response Model}

When calibration is used to estimate the response model, it is required to know the population total $\sum_{k \in U}\zb_k$. 
By slightly modifying the estimating equation, calibration can nevertheless be applied when no information is available on population units outside the sample. Specifically, replacing the population totals in the first term of Equations~\eqref{estimating:eqn:calib} with their corresponding HT estimators yields
\begin{align} \label{estimating:eqn:calib2}
   \sum_{k \in s} \frac{\zb_k}{\pi_k} - \sum_{k \in s_r} \frac{\zb_k}{\pi_k}F(\zb_k^\top\lambdab) &= 0.
\end{align}
This modification enables estimation even when no auxiliary information is known for units outside the sample. Alternative approaches in this setting include maximum likelihood and local polynomial regression (see Sections~\ref{section:mle} and \ref{section:proba:localpoly}).

%Other choices to estimate the response probabilities when the values $\{\xb_k\}$ are only available for the sampled units are maximum likelihood and local polynomial regression (see Sections~\ref{section:mle} and \ref{section:proba:localpoly}, respectively).  
%Depending on the level of knowledge of the auxiliary variables and the selection of working and response models, various combinations of estimators (e.g., $\widehat{t}_{qma}$, $\widehat{t}_{qma,\pi}$) and methods for estimating response probabilities (such as calibration on true totals, calibration on HT estimators, maximum likelihood, local polynomial regression, etc.) are possible.
%This goes beyond the scope of our work. 
%Common choices for the weights are 1 or $\pi_k^{-1}$. When $c_k=1$ usual maximum likelihood estimation is applied. 

% ----------------------------------------------------------------
\subsection{Examples of Existing Estimators in the Class of Quasi-Model-Assisted Estimators}\label{section:existing:DR}
% ----------------------------------------------------------------

The class of quasi-model-assisted estimators is broad and encompasses many existing estimators. This section presents some examples, beginning with three estimators in the survey sampling context.

\begin{example}[Classical NWA estimator with Calibrated Response Probabilities]
Suppose that the same auxiliary variables are used in both the working model and the response model. In this case, when the response probabilities are estimated by calibration as presented in Section~\ref{section:calibration}, and the working model is the linear model described in \textit{Section~\ref{section:linear:working:model}, the quasi-model-assisted estimator reduces to the classical NWA estimator with response probabilities estimated via calibration. Indeed, in this setting, the quasi-model-assisted estimator is
 \begin{align}
\widehat{t}_{qma} &= \sum_{k \in s_r} \frac{y_k}{\pi_k \widehat{p}_k} + \left(\sum_{k \in U} \xb_k - \sum_{k \in s_r} \frac{\xb_k}{\pi_k \widehat{p}_k}\right)^\top\widehat{\Bb}_r = \sum_{k \in s_r} \frac{y_k}{\pi_k \widehat{p}_k},
\end{align}
where the last equality follows from the calibration Equation~\eqref{estimating:eqn:calib}. This results holds for any choice of $F(\cdot)$.}
\end{example}

\begin{comment}
\begin{example}[GREG Estimator On The Respondents]
When estimation of the response probabilities via calibration with linear weight adjustment $F(u) = 1+u$ and the linear working model. If we set $c_k = \pi_k$ in \eqref{equation:beta} and consider the same variables in both working and response models, i.e., $\xb_k = \zb_k$, the quasi-model-assisted estimator in \eqref{estimator:diff:mr:U:phat} can be written as
\begin{align}
	\widehat{t}_{qma} &= \sum_{k \in s_r} \dfrac{y_k}{\pi_k} + 
        \left(\sum_{k \in U} \xb_k  - \sum_{k \in s_r} \frac{\xb_k}{\pi_k} \right) ^\top  \widehat{\mathbf{B}}_r.
\end{align}
This estimator is particular case of GREG estimator where the units $k \in s_r$ are considered instead of the sampled unit $k \in s$.     
\end{example}
\end{comment}

\begin{example}[Imputed Estimator of \cite{bea:05a}]
\cite{bea:05a} proposes an imputation procedure yielding an imputed estimator that coincides with a quasi-model-assisted estimator. 
When the parameter of interest is the population total, the quasi-model-assisted estimator considered in \cite{bea:05a} is
\begin{align}\label{equation:calib:imputation:beaumont}
    \sum_{k \in s} \widetilde{w}_k m(\xb_k;\widehat{\Bb}) + \sum_{k \in s_r} \frac{\widetilde{w}_k}{p_k(\widehat{\lambdab})}\left\{y_k - m(\xb_k;\widehat{\Bb})\right\},
\end{align}
where $\widetilde{w}_k = g_k/\pi_k$ are calibrated weights and both the working model $m(\xb_k;\betab)$ and the response model $p_k(\lambdab)$ are parametric. The estimated parameter vectors are $\widehat{\Bb}$ and $\widehat{\lambdab}$, respectively. This estimator is a special case of the quasi–model-assisted class, with two differences as compared with the general form of quasi-model-assisted estimators as defined in Equation~\eqref{estimator:diff:mr:U:phat}: (i) the first term is summed over the sample rather than using the population total as in Section~\ref{section:qmapi}, and (ii) calibration-adjusted weights $\widetilde{w}_k$ replace the design weights. 
%A parametric working model $\E(y_k|\xb_k)=m(\xb_k;\betab)$ is assumed, where $m$ is a known function and $\widehat{\Bb}$ an estimator such as the maximum likelihood estimator of the unknown vector of parameter $\betab$. A parametric response model $p_k(\lambdab)$ is assumed, where $p_k(\cdot)$ is a known function of an unknown vector of parameters, and $\widehat{\lambdab}$ is an estimator of $\lambda$. The estimator in Equation~\eqref{equation:calib:imputation:beaumont} belongs to the general class of quasi-model-assisted estimators with both the working model and the response model assumed to be parametric. Two small differences with the estimator in Equation~\ref{estimator:diff:mr:U:phat} are to be noted. First, the first term of the estimator is a reweighted estimator with a sum over the sampled units. The population sum is considered in the estimator in Equation~\ref{estimator:diff:mr:U:phat}. See also Section~\ref{section:qmapi} for more details. Second, calibrated weights $\widetilde{w}_k$ are used instead of the design weights. 
\end{example}

\begin{example}[Doubly Robust Estimator of \cite{kim:haz:14}]
In the survey sampling context, \cite{kim:haz:14} present a class of estimator
\begin{align}\label{equation:dr:estimator:kim:haziza}
    \sum_{k \in s} \frac{m(\xb_k;\widehat{\Bb})}{\pi_k}  + \sum_{k \in s_r} \frac{y_k - m(\xb_k;\widehat{\Bb})}{\pi_k p_k(\widehat{\lambdab})}.
\end{align}
The working model $\E(y_k|\xb_k)=m(\xb_k;\betab)$ and the response model $p_k(\lambdab)$ are both parametric. The estimated parameters $\widehat{\Bb}$ and $\widehat{\lambdab}$ are the solutions to a system of estimating equations. With these solutions, the resulting point estimator in~\eqref{equation:dr:estimator:kim:haziza} and its variance estimator are doubly robust.
\end{example}

Some estimators used outside the survey sampling framework can also be expressed as quasi-model-assisted estimators. We present two examples below. In survey sampling, the parameter of interest is typically a finite-population quantity, such as the total $\sum_{k \in U}y_k$ or mean $N^{-1}\sum_{k \in U}y_k$. In the two examples below, however, the parameter of interest is an infinite-population quantity, such as the unconditional mean of $y$.

\begin{example}[Augmented Inverse Probability Weighted Estimator (AIPW)] \label{example:aipw}
    The AIPW estimator is widely used for causal inference, particularly to estimate average treatment effects from observational data \cite{rob:rot:zha:94,rot:rob:sch:98,ban:rob:05}. Let $y$ be an outcome variable that may be missing for some subjects, and let $\xb$ be a vector of auxiliary variables observed for all subjects. Denote by $\delta$ the indicator taking value 1 if $y$ is observed and 0 otherwise. Suppose that the parameter of interest is the unconditional mean $\mu$ of $y$ based on $N$ i.i.d. copies of the observed data $(\delta_k, d_{\mbox{obs},k})$, $k = 1,\ldots,N$, where $d_{\mbox{obs},k} = (\xb_k^\top, y_k)^\top$ if $\delta_k = 1$ and $\xb_k$ if $\delta_k = 0$.
    Given the observed data, one can model the conditional mean $\E(y \mid \delta=1, \xb)$ and the probability of observation $\Pr(\delta=1 \mid \xb)$ using parametric forms $m(\xb;\betab)$ and $p(\xb;\lambdab)$, with unknown parameters $\betab$ and $\lambdab$. The AIPW estimator is then
    \begin{align}
    \widehat{\mu}_{AIPW} = \frac{1}{N} \sum_{k=1}^{N} \Big[ m(\xb_k; \widehat{\betab}) + \frac{\delta_k}{p(\xb_k, \widehat{\lambdab})} \big\{y_k - m(\xb_k; \widehat{\betab})\big\} \Big],
    \end{align}
    where $\widehat{\betab}$ and $\widehat{\lambdab}$ solve estimating equations ensuring double robustness.  
    The AIPW estimator can be seen as a particular case of a quasi–model-assisted estimator with $\pi_k = 1$ for all $k$, i.e., a full census, and parametric working and response models. The AIPW is not suitable when data arise from a complex sampling design.
\end{example}

\begin{example}[Regression Estimation with Residual Bias Correction]
%\cite{kan:sch:07} present three procedures to construct doubly robust estimators. One of these procedure is regression estimation with residual bias correction. This procedure can be described as follows. Consider the same notation as in Example~\ref{example:aipw}. Suppose that a score function $U_k$ for unit $k$ is such that the estimator of $\mu$ in the case of complete response is the solution to estimating equation $\sum_{k = 1}^N U_k = 0$. For instance, the sample mean $N^{-1}\sum_{k = 1}^N$ is the solution to this estimating equation with $U_k = (y_k - \mu)/\sigma^2$ for any $\sigma^2 > 0$. 
%\cite{kan:sch:07} present the estimating equation
%\begin{align}\label{equation:drU}
%\frac{1}{N}\sum_{i = 1}^N \widehat{U}_k + \frac{1}{N}\sum_{i = 1}^N \delta_k \widehat{p}_k^{-1} (U_k - \widehat{U}_k) = 0,      
% \end{align}
% where $\widehat{U}_k$ is a quasi-score function with $y_k$ replaced in $U_k$ with an estimate of the regression function $m(\xb_k)$. The solution to estimating Equation~\ref{equation:drU} is a quasi-model-assisted estimator and an AIPW estimator when $U_k = (y_k - \mu)/\sigma^2$, $\widehat{m}(\xb_k) = \widehat{m}_r(\xb_k)$, $\delta_k = r_k$ is the membership indicator of the set of respondents, and $\pi_k = 1$ for all $k$. Variations of estimating Equation~\ref{equation:drU}, such as normalizing the weights, are possible. See \cite{kan:sch:07} for more details.  
\cite{kan:sch:07} propose regression estimation with residual bias correction as a doubly robust procedure. 
Consider the same notation as in Example~\ref{example:aipw}.
Let $U_k$ be a score function such that $\sum_{k=1}^N U_k = 0$ gives the complete-data estimator of $\mu$. The estimating equation is
\begin{align}\label{equation:drU}
\frac{1}{N}\sum_{k=1}^N \widehat{U}_k + \frac{1}{N}\sum_{k=1}^N \delta_k \widehat{p}_k^{-1} (U_k - \widehat{U}_k) = 0,
\end{align}
where $\widehat{U}_k$ is the analogous of $U_k$ where $y_k$ is replaced by an estimate of $m(\xb_k)$. Its solution is a quasi–model-assisted estimator and reduces to AIPW when $U_k = (y_k-\mu)/\sigma^2$, $\widehat{m}(\xb_k) = \widehat{m}_r(\xb_k)$, $\delta_k = r_k$, and $\pi_k = 1$. Variants, such as weight normalization, are possible, see \cite{kan:sch:07}.
\end{example}

% ----------------------------------------------------------------
\section{Simulations}\label{section:simulation}
% ----------------------------------------------------------------

Let $U$ be a population of size $N=2000$. For each unit $k \in U$, a vector of auxiliary variables $\xb_k = (x_{k1}, x_{k2})^\top$ is generated from independent and identically distributed random uniform variables with parameters -5 and 5. The survey variable $y$ is generated as
$$
y_k = 115 + 10.4\, x_{k1} + 9.7\, x_{k2} + \varepsilon_k,
$$
where $\varepsilon_k$ is the realization of a normal distribution of mean 0 and standard deviation 15.
The value $y_k$ is observed with probability
\begin{align}
p_k = \big\{ 1+\exp[-\lambdab^\top (1, x_{k1}, x_{k2})^\top] \big\}^{-1}, \quad \lambdab = (0.010, -0.198, 0.202)^\top,
\end{align}
and missing otherwise. With these specifications, approximately $50\%$ of the values are missing.

%\re{There is no clear working model define}
We consider different scenarios, depending on whether the response and working models are well specified or misspecified.

To study model misspecification, we define alternative auxiliary variables
\begin{align}
\widetilde{x}_{k1} &= 10 \cos(x_{k2}) \cdot \big\{ 1 + \exp(x_{k1}) \big\}^{-1} + 10,\\
\widetilde{x}_{k2} &= 0.2 \cdot \big( \sqrt{|x_{k2}-x_{k1}|} + 10 \big) \cdot |x_{k1}|.
\end{align}
The correlation between the different considered variables is reported in Table~\ref{tablecor}. The vector $\xb_k$ is strongly related to both $y_k$ and $p_k$, while $\widetilde{\xb}_k$ is weakly related and is used to misspecify the models.

\begin{table}[htb]
  \caption{Correlation of the variable of interest $\{ y_k \}$ and the response probabilities $\{ p_k \}$ with the auxiliary variables  $\{x_{k1}\}, \{x_{k2}\}, \{\widetilde{x}_{k1}\}$, and $\{\widetilde{x}_{k2}\}$, for population values $k \in U$. \label{tablecor}}
  \centering
  \resizebox{7cm}{!}{
  \fontsize{9}{11}\selectfont
  \begin{tabular}[t]{c|cccc}
  \toprule
    & $\{x_{k1}\}$ & $\{x_{k2}\}$ & $\{\widetilde{x}_{k1}\}$ & $\{\widetilde{x}_{k2}\}$\\
   \hline
   $\{ y_k \}$  & 0.688 & 0.647 & 0.127 & 0.026 \\
   $\{ p_k \}$  & 0.672 & 0.742 & 0.112 & 0.021 \\
  \bottomrule
  \end{tabular}}
\end{table}

Four scenarios are considered with the response and working model fitted with different pairs of auxiliary variables. Table~\ref{table1} indicates which pair of auxiliary variables is used to fit the models in all four scenarios. Scenarios 1 and 2 have the working model well specified, while 3 and 4 have it misspecified. 
Scenarios 1 and 3 have the response model well specified, while 2 and 4 have it misspecified.  Thus, both models are well specified in scenario 1, only one model is well specified in scenarios 2 and 3, and neither model is well specified in scenario 4.
\begin{table}[htb!]
  \caption{Pair of auxiliary variable used to estimate the response probabilities ${ \widehat{p}_k }$ and the function $\widehat{m}_r(.)$ across four scenarios.  \label{table1}}
  \centering
  \resizebox{8cm}{!}{
  \fontsize{9}{11}\selectfont
  \begin{tabular}[t]{cc|cc}
  \toprule
   & & \multicolumn{2}{c}{$\widehat{p}_k,~ \widehat{p}_{k,\pi}$} \\
  &  & $\{x_{k1}, x_{k2}\}$ & $\{\widetilde{x}_{k1}, \widetilde{x}_{k2}\}$ \\
  \midrule
  \multirow{2}{3em}{$\widehat{m}_r(.)$} &  $\{x_{k1}, x_{k2}\}$ & Scenario 1 & Scenario 2 \\
  &$\{\widetilde{x}_{k1}, \widetilde{x}_{k2}\}$& Scenario 3 & Scenario 4 \\
  \bottomrule
  \end{tabular}}
\end{table}

% We compare four estimators: $\widehat{t}_{NWA}$ and $\widehat{t}_{qma,\pi}$, defined in Sections~\ref{section:NWA} and \ref{section:qmapi}, respectively, the imputed estimator $\widehat{t}_{imp} = \sum_{k\in s_r} y_k/\pi_k + \sum_{k\in s\backslash s_r} m_r(\xb_k)/\pi_k$, and the naive estimator $\widehat{t}_{naive}= N n_r^{-1} \cdot \sum_{k\in s_r} y_k$.
% Note that these estimators only require the covariates to be known at the sample level.
% Estimators $\widehat{t}_{imp}$ and $\widehat{t}_{qma,\pi}$ depend on the estimated function $\widehat{m}_r(.)$. 
% Four different prediction methods are used to obtain $\widehat{m}_r(.)$: generalized linear regression, local polynomial regression, $K$-nearest neighbors with $K=5$ and random forest. 
% Estimators $\widehat{t}_{NWA}$ and $\widehat{t}_{qma,\pi}$ depend on the estimated response probabilities $\{ \widehat{p}_{k} \}$.

We compare four estimators: the NWA estimator $\widehat{t}_{NWA}$ defined in Equation~\eqref{estimatornwa}, the quasi-model-assisted estimator $\widehat{t}_{qma,\pi}$ defined in Equation~\eqref{estimator:diff:mr:U:phat:pi}, the imputed estimator
$$
\widehat{t}_{imp} = \sum_{k\in s_r} \frac{y_k}{\pi_k} + \sum_{k\in s \setminus s_r} \frac{\widehat{m}_r(\xb_k)}{\pi_k},
$$
and the naive estimator $\widehat{t}_{naive} = N n_r^{-1} \sum_{k \in s_r} y_k$.  

The imputed estimator $\widehat{t}_{imp}$ depends on the estimated regression function $\widehat{m}_r$, which is obtained using four prediction methods: generalized linear regression, local polynomial regression, $K$-nearest neighbors ($K=5$), and random forest. 
The NWA estimator $\widehat{t}_{NWA}$ depends on the estimated response probabilities $\{\widehat{p}_k\}$. Three methods are used to estimate the response probabilities: calibration based on Equation~\eqref{estimating:eqn:calib2} using a logistic model, maximum likelihood estimation via Equation~\eqref{eqn:mle:logistic} with $c_k = 1$, and the $K$-nearest neighbors ($K = 10$). 
The quasi-model-assisted estimator $\widehat{t}_{qma,\pi}$ depends on the estimated regression function $\widehat{m}_r(\cdot)$, and on the estimated response probabilities $\{\widehat{p}_k\}$. We compute the estimator $\widehat{t}_{qma,\pi}$ for all possible 12 combinations of estimation methods discussed above. Details on the implementation of the methods are provided in Appendix~\ref{appendix:detailofthemodel}. 

With these choices, the imputed estimators $\widehat{t}_{imp}$, the NWA estimators $\widehat{t}_{NWA}$, and the quasi-model-assisted estimators $\widehat{t}_{qma,\pi}$ all require the same level of auxiliary information. Indeed, these estimators require the auxiliary variables to be known at the sample level. No auxiliary information is needed for the units outside the sample. However, the naive estimator require no auxiliary information.

We select $I=10'000$ samples of expected size $n=200$ from the population $U$ using a Poisson sampling design with inclusion probabilities 
$\pi_k = 0.1$ if $x_{k1}<0$ and $\pi_k = 0.3$ otherwise.  In what follows, the superscript $(i)$ denotes quantities associated with simulation run $i, i = 1, \ldots, I$.
For each selected samples $s^{(i)}$, missing values in the survey variable are generated at random according to response probabilities $\{ p_k \}$ using a Poisson sampling design.
The expected number of observed values in each sample, denoted $n_r^{(i)}$, is $n/2 = 100$.
Let $s^{(i)}_r \subset s^{(i)}$ denote the sub-sample of size $n_r^{(i)}$ consisting of units for which $y_k$ is observed at simulation run~$i$.

To evaluate the quality of the estimated response and working models, two quantities are computed: the average over all simulation runs of the Mean Absolute Error (MAE) of the estimated response probabilities $\{ \widehat{p}_k \}$
$$
  \mbox{MAE}(\widehat{p}_k) = \frac{1}{I}\sum_{i = 1}^I\frac{1}{n_r^{(i)}} \sum_{k\in s_r^{(i)}} \mid \widehat{p}_k^{(i)} - p_k \mid,
$$
and of the Mean Relative Prediction Error (MRPE)
$$
  \mbox{MRPE}(\widehat{m}_r(\ub_k)) = \frac{1}{I} \sum_{i = 1}^I \dfrac{N^{-1} \cdot \sum_{k\in U} \mid \widehat{m}_r^{(i)}(\ub_k) - y_k \mid}{\sum_{k\in U} y_k},
$$
where $\ub_k = \xb_k$ or $\ub_k = \widetilde{\xb}_k$, depending on the scenario. 
Table~\ref{table2} reports these averages. The response model shows a better fit in scenarios 1 and 3 than in scenarios 2 and 4, with MAE values between 0.03 and 0.11 in scenarios 1 and 3, and between 0.16 and 0.19 in scenarios 2 and 4. The working model shows a better fit in scenarios 1 and 2 than in scenarios 3 and 4, with MRPE values between 0.10 and 0.12 in scenarios 1 and 2, and between 0.26 and 0.37 in scenarios 3 and 4.
A more detailed discussion of the figures shown in Table~\ref{table2} is provided in Appendix~\ref{appendix:model:quality}.

% \begin{table}[!ht]
%     \centering
%     \caption{MAE of the estimated response probabilities $\widehat{p}_k$ and $\widehat{p}_{k,\pi}$ and MRPE of $\widehat{m}_r(\ub_k)$ of the prediction of the missing values over all simulation runs in four scenarios for the simulated data. \label{table2}}
%     \centering
%     \resizebox{8cm}{!}{
%     \fontsize{9}{11}\selectfont
%     \begin{tabular}[t]{lrrrr}
%         \toprule
%         \multicolumn{1}{c}{ } & \multicolumn{4}{c}{Scenario} \\
%         \cmidrule(l{3pt}r{3pt}){2-5}
%           & 1 & 2 & 3 & 4\\
%         \midrule
%         $\mbox{MAE} (\widehat{p}_k)$ & 0.078 & 0.194 & 0.078 & 0.194\\
%         $\mbox{MAE} (\widehat{p}_{k,\pi})$
%          & 0.060 & 0.188 & 0.060 & 0.188\\
%         \addlinespace[1ex]
%         \multicolumn{5}{l}{$\mbox{MRPE}$}\\
%         \hspace{1em}GREG & 0.110 & 0.110 & 0.374 & 0.374\\
%         \hspace{1em}poly & 0.117 & 0.117 & 0.320 & 0.320\\
%         \hspace{1em}$K$-nn & 0.128 & 0.128 & 0.354 & 0.354\\
%         \bottomrule
%     \end{tabular}}
% \end{table}

\begin{table}[!ht]
    \centering
    \caption{MAE of estimated response probabilities $\widehat{p}_k$ for three estimation methods, and MRPE of $\widehat{m}_r(\mathbf{u}_k)$ for four prediction methods, computed over all simulation runs across the four simulated scenarios. \label{table2}}
    \centering
    \resizebox{12cm}{!}{
    \fontsize{9}{11}\selectfont
    \begin{tabular}[t]{lrrrr}
        \toprule
         Scenario & 1 & 2 & 3 & 4\\
        \midrule
        $\mbox{MAE}(\widehat{p}_k)$ & $\widehat{p}_k $ \checkmark & $\widehat{p}_k $ \(\times\) & $\widehat{p}_k $ \checkmark & $\widehat{p}_k $ \(\times\)\\
         \hspace{1em} Calibration & 0.037 & 0.180  & 0.037 & 0.180 \\
         \hspace{1em} Maximum Likelihood
         & 0.032  &  0.186 & 0.032 & 0.186 \\
          \hspace{1em} $K$-nearest neighbor 
         & 0.113  & 0.164  & 0.113 & 0.164\\
        \addlinespace[1ex]
        $\mbox{MRPE}(\widehat{m}_r(\ub_k))$ & $\widehat{m}_r ( \ub_k )$ \checkmark & $\widehat{m}_r ( \ub_k )$ \checkmark & $\widehat{m}_r ( \ub_k )$ \(\times\) & $\widehat{m}_r ( \ub_k )$ \(\times\)\\
        \hspace{1em} Linear & 0.104  & 0.104  & 0.366 & 0.366 \\
        \hspace{1em} Polynomial & 0.107  & 0.107  & 0.287 & 0.287 \\
        \hspace{1em}$K$-nearest neighbor & 0.121  & 0.121  & 0.283 & 0.283 \\
        \hspace{1em}Random forest & 0.119  & 0.119 & 0.258 & 0.258 \\
        \bottomrule
    \end{tabular}}
\end{table}

For each samples $s^{(i)}$ or $s_r^{(i)}$, we estimate the population total with the aforementioned total estimators.
For a generic total estimator $\widehat{t}$, we compute the Monte Carlo bias relative to the true total
$$
  \mbox{RB}(\widehat{t}) = \dfrac{I^{-1} \sum_{i=1}^I (\widehat{t}^{(i)} - t)}{t}
$$
and the Monte Carlo standard deviation relative to the true total
$$
  \mbox{RSd}(\widehat{t}) =  \dfrac{  \sqrt{ (I-1)^{-1} \sum_{i=1}^I (\widehat{t}^{(i)} - t)^2}}{t}.
$$
We compare the total estimators for each scenario. Figure~\ref{plotres} summarizes the results and the detailed results are given in Appendix~\ref{qma:appendix:D} in Tables~\ref{table1:simulateddata} and~\ref{table2:simulateddata}.

\begin{figure}[htb!]
	\centering
        \resizebox{14cm}{!}{
% Created by tikzDevice version 0.12.6 on 2025-11-12 17:08:01
% !TEX encoding = UTF-8 Unicode
% Created by tikzDevice version 0.12.6 on 2025-11-13 16:51:13
% !TEX encoding = UTF-8 Unicode
\begin{tikzpicture}[x=1pt,y=1pt]
\definecolor{fillColor}{RGB}{255,255,255}
\path[use as bounding box,fill=fillColor,fill opacity=0.00] (0,0) rectangle (469.85,361.35);
\begin{scope}
\path[clip] (  0.00,  0.00) rectangle (469.85,361.35);
\definecolor{drawColor}{RGB}{255,255,255}
\definecolor{fillColor}{RGB}{255,255,255}

\path[draw=drawColor,line width= 0.6pt,line join=round,line cap=round,fill=fillColor] (  0.00,  0.00) rectangle (469.85,361.35);
\end{scope}
\begin{scope}
\path[clip] ( 38.56,187.73) rectangle (136.74,339.28);
\definecolor{fillColor}{RGB}{255,255,255}

\path[fill=fillColor] ( 38.56,187.73) rectangle (136.74,339.28);
\definecolor{drawColor}{gray}{0.87}

\path[draw=drawColor,line width= 0.1pt,line join=round] ( 38.56,211.11) --
	(136.74,211.11);

\path[draw=drawColor,line width= 0.1pt,line join=round] ( 38.56,240.54) --
	(136.74,240.54);

\path[draw=drawColor,line width= 0.1pt,line join=round] ( 38.56,269.98) --
	(136.74,269.98);

\path[draw=drawColor,line width= 0.1pt,line join=round] ( 38.56,299.42) --
	(136.74,299.42);

\path[draw=drawColor,line width= 0.1pt,line join=round] ( 38.56,328.86) --
	(136.74,328.86);

\path[draw=drawColor,line width= 0.3pt,line join=round] ( 38.56,196.39) --
	(136.74,196.39);

\path[draw=drawColor,line width= 0.3pt,line join=round] ( 38.56,225.82) --
	(136.74,225.82);

\path[draw=drawColor,line width= 0.3pt,line join=round] ( 38.56,255.26) --
	(136.74,255.26);

\path[draw=drawColor,line width= 0.3pt,line join=round] ( 38.56,284.70) --
	(136.74,284.70);

\path[draw=drawColor,line width= 0.3pt,line join=round] ( 38.56,314.14) --
	(136.74,314.14);

\path[draw=drawColor,line width= 0.3pt,line join=round] ( 52.58,187.73) --
	( 52.58,339.28);

\path[draw=drawColor,line width= 0.3pt,line join=round] ( 75.96,187.73) --
	( 75.96,339.28);

\path[draw=drawColor,line width= 0.3pt,line join=round] ( 99.33,187.73) --
	( 99.33,339.28);

\path[draw=drawColor,line width= 0.3pt,line join=round] (122.71,187.73) --
	(122.71,339.28);
\definecolor{drawColor}{RGB}{0,0,0}
\definecolor{fillColor}{RGB}{0,0,0}

\path[draw=drawColor,line width= 0.4pt,line join=round,line cap=round,fill=fillColor] ( 52.58,331.80) circle (  2.50);

\path[draw=drawColor,line width= 0.4pt,line join=round,line cap=round,fill=fillColor] ( 99.33,196.39) circle (  2.50);

\path[draw=drawColor,line width= 0.4pt,line join=round,line cap=round,fill=fillColor] ( 99.33,194.62) circle (  2.50);

\path[draw=drawColor,line width= 0.4pt,line join=round,line cap=round,fill=fillColor] ( 99.33,195.80) circle (  2.50);

\path[draw=drawColor,line width= 0.4pt,line join=round,line cap=round,fill=fillColor] ( 75.96,195.21) circle (  2.50);

\path[draw=drawColor,line width= 0.4pt,line join=round,line cap=round,fill=fillColor] (122.71,196.39) circle (  2.50);

\path[draw=drawColor,line width= 0.4pt,line join=round,line cap=round,fill=fillColor] (122.71,196.39) circle (  2.50);

\path[draw=drawColor,line width= 0.4pt,line join=round,line cap=round,fill=fillColor] (122.71,196.39) circle (  2.50);

\path[draw=drawColor,line width= 0.4pt,line join=round,line cap=round,fill=fillColor] ( 75.96,195.80) circle (  2.50);

\path[draw=drawColor,line width= 0.4pt,line join=round,line cap=round,fill=fillColor] ( 75.96,206.98) circle (  2.50);

\path[draw=drawColor,line width= 0.4pt,line join=round,line cap=round,fill=fillColor] (122.71,199.33) circle (  2.50);

\path[draw=drawColor,line width= 0.4pt,line join=round,line cap=round,fill=fillColor] (122.71,199.33) circle (  2.50);

\path[draw=drawColor,line width= 0.4pt,line join=round,line cap=round,fill=fillColor] (122.71,199.33) circle (  2.50);

\path[draw=drawColor,line width= 0.4pt,line join=round,line cap=round,fill=fillColor] ( 75.96,211.69) circle (  2.50);

\path[draw=drawColor,line width= 0.4pt,line join=round,line cap=round,fill=fillColor] (122.71,204.63) circle (  2.50);

\path[draw=drawColor,line width= 0.4pt,line join=round,line cap=round,fill=fillColor] (122.71,204.63) circle (  2.50);

\path[draw=drawColor,line width= 0.4pt,line join=round,line cap=round,fill=fillColor] (122.71,204.63) circle (  2.50);

\path[draw=drawColor,line width= 0.6pt,dash pattern=on 4pt off 4pt ,line join=round] ( 38.56,196.39) -- (136.74,196.39);
\definecolor{drawColor}{gray}{0.70}

\path[draw=drawColor,line width= 0.6pt,line join=round,line cap=round] ( 38.56,187.73) rectangle (136.74,339.28);
\end{scope}
\begin{scope}
\path[clip] ( 38.56, 30.69) rectangle (136.74,182.23);
\definecolor{fillColor}{RGB}{255,255,255}

\path[fill=fillColor] ( 38.56, 30.69) rectangle (136.74,182.23);
\definecolor{drawColor}{gray}{0.87}

\path[draw=drawColor,line width= 0.1pt,line join=round] ( 38.56, 54.06) --
	(136.74, 54.06);

\path[draw=drawColor,line width= 0.1pt,line join=round] ( 38.56, 83.50) --
	(136.74, 83.50);

\path[draw=drawColor,line width= 0.1pt,line join=round] ( 38.56,112.94) --
	(136.74,112.94);

\path[draw=drawColor,line width= 0.1pt,line join=round] ( 38.56,142.37) --
	(136.74,142.37);

\path[draw=drawColor,line width= 0.1pt,line join=round] ( 38.56,171.81) --
	(136.74,171.81);

\path[draw=drawColor,line width= 0.3pt,line join=round] ( 38.56, 39.34) --
	(136.74, 39.34);

\path[draw=drawColor,line width= 0.3pt,line join=round] ( 38.56, 68.78) --
	(136.74, 68.78);

\path[draw=drawColor,line width= 0.3pt,line join=round] ( 38.56, 98.22) --
	(136.74, 98.22);

\path[draw=drawColor,line width= 0.3pt,line join=round] ( 38.56,127.65) --
	(136.74,127.65);

\path[draw=drawColor,line width= 0.3pt,line join=round] ( 38.56,157.09) --
	(136.74,157.09);

\path[draw=drawColor,line width= 0.3pt,line join=round] ( 52.58, 30.69) --
	( 52.58,182.23);

\path[draw=drawColor,line width= 0.3pt,line join=round] ( 75.96, 30.69) --
	( 75.96,182.23);

\path[draw=drawColor,line width= 0.3pt,line join=round] ( 99.33, 30.69) --
	( 99.33,182.23);

\path[draw=drawColor,line width= 0.3pt,line join=round] (122.71, 30.69) --
	(122.71,182.23);
\definecolor{drawColor}{RGB}{0,0,0}
\definecolor{fillColor}{RGB}{0,0,0}

\path[draw=drawColor,line width= 0.4pt,line join=round,line cap=round,fill=fillColor] ( 52.58,175.34) circle (  2.50);

\path[draw=drawColor,line width= 0.4pt,line join=round,line cap=round,fill=fillColor] ( 99.33, 46.41) circle (  2.50);

\path[draw=drawColor,line width= 0.4pt,line join=round,line cap=round,fill=fillColor] ( 99.33, 69.96) circle (  2.50);

\path[draw=drawColor,line width= 0.4pt,line join=round,line cap=round,fill=fillColor] ( 99.33, 74.67) circle (  2.50);

\path[draw=drawColor,line width= 0.4pt,line join=round,line cap=round,fill=fillColor] ( 75.96, 69.37) circle (  2.50);

\path[draw=drawColor,line width= 0.4pt,line join=round,line cap=round,fill=fillColor] (122.71, 69.37) circle (  2.50);

\path[draw=drawColor,line width= 0.4pt,line join=round,line cap=round,fill=fillColor] (122.71, 69.37) circle (  2.50);

\path[draw=drawColor,line width= 0.4pt,line join=round,line cap=round,fill=fillColor] (122.71, 69.37) circle (  2.50);

\path[draw=drawColor,line width= 0.4pt,line join=round,line cap=round,fill=fillColor] ( 75.96, 71.72) circle (  2.50);

\path[draw=drawColor,line width= 0.4pt,line join=round,line cap=round,fill=fillColor] (122.71, 69.96) circle (  2.50);

\path[draw=drawColor,line width= 0.4pt,line join=round,line cap=round,fill=fillColor] (122.71, 69.96) circle (  2.50);

\path[draw=drawColor,line width= 0.4pt,line join=round,line cap=round,fill=fillColor] (122.71, 69.96) circle (  2.50);

\path[draw=drawColor,line width= 0.4pt,line join=round,line cap=round,fill=fillColor] ( 75.96, 73.49) circle (  2.50);

\path[draw=drawColor,line width= 0.4pt,line join=round,line cap=round,fill=fillColor] (122.71, 71.13) circle (  2.50);

\path[draw=drawColor,line width= 0.4pt,line join=round,line cap=round,fill=fillColor] (122.71, 71.13) circle (  2.50);

\path[draw=drawColor,line width= 0.4pt,line join=round,line cap=round,fill=fillColor] (122.71, 71.13) circle (  2.50);

\path[draw=drawColor,line width= 0.6pt,dash pattern=on 4pt off 4pt ,line join=round] ( 38.56, 39.34) -- (136.74, 39.34);
\definecolor{drawColor}{gray}{0.70}

\path[draw=drawColor,line width= 0.6pt,line join=round,line cap=round] ( 38.56, 30.69) rectangle (136.74,182.23);
\end{scope}
\begin{scope}
\path[clip] (142.24,187.73) rectangle (240.42,339.28);
\definecolor{fillColor}{RGB}{255,255,255}

\path[fill=fillColor] (142.24,187.73) rectangle (240.42,339.28);
\definecolor{drawColor}{gray}{0.87}

\path[draw=drawColor,line width= 0.1pt,line join=round] (142.24,211.11) --
	(240.42,211.11);

\path[draw=drawColor,line width= 0.1pt,line join=round] (142.24,240.54) --
	(240.42,240.54);

\path[draw=drawColor,line width= 0.1pt,line join=round] (142.24,269.98) --
	(240.42,269.98);

\path[draw=drawColor,line width= 0.1pt,line join=round] (142.24,299.42) --
	(240.42,299.42);

\path[draw=drawColor,line width= 0.1pt,line join=round] (142.24,328.86) --
	(240.42,328.86);

\path[draw=drawColor,line width= 0.3pt,line join=round] (142.24,196.39) --
	(240.42,196.39);

\path[draw=drawColor,line width= 0.3pt,line join=round] (142.24,225.82) --
	(240.42,225.82);

\path[draw=drawColor,line width= 0.3pt,line join=round] (142.24,255.26) --
	(240.42,255.26);

\path[draw=drawColor,line width= 0.3pt,line join=round] (142.24,284.70) --
	(240.42,284.70);

\path[draw=drawColor,line width= 0.3pt,line join=round] (142.24,314.14) --
	(240.42,314.14);

\path[draw=drawColor,line width= 0.3pt,line join=round] (156.26,187.73) --
	(156.26,339.28);

\path[draw=drawColor,line width= 0.3pt,line join=round] (179.64,187.73) --
	(179.64,339.28);

\path[draw=drawColor,line width= 0.3pt,line join=round] (203.01,187.73) --
	(203.01,339.28);

\path[draw=drawColor,line width= 0.3pt,line join=round] (226.39,187.73) --
	(226.39,339.28);
\definecolor{drawColor}{RGB}{0,0,0}
\definecolor{fillColor}{RGB}{0,0,0}

\path[draw=drawColor,line width= 0.4pt,line join=round,line cap=round,fill=fillColor] (156.26,331.80) circle (  2.50);

\path[draw=drawColor,line width= 0.4pt,line join=round,line cap=round,fill=fillColor] (203.01,306.48) circle (  2.50);

\path[draw=drawColor,line width= 0.4pt,line join=round,line cap=round,fill=fillColor] (203.01,317.67) circle (  2.50);

\path[draw=drawColor,line width= 0.4pt,line join=round,line cap=round,fill=fillColor] (203.01,222.29) circle (  2.50);

\path[draw=drawColor,line width= 0.4pt,line join=round,line cap=round,fill=fillColor] (179.64,195.21) circle (  2.50);

\path[draw=drawColor,line width= 0.4pt,line join=round,line cap=round,fill=fillColor] (226.39,195.80) circle (  2.50);

\path[draw=drawColor,line width= 0.4pt,line join=round,line cap=round,fill=fillColor] (226.39,195.80) circle (  2.50);

\path[draw=drawColor,line width= 0.4pt,line join=round,line cap=round,fill=fillColor] (226.39,195.80) circle (  2.50);

\path[draw=drawColor,line width= 0.4pt,line join=round,line cap=round,fill=fillColor] (179.64,195.80) circle (  2.50);

\path[draw=drawColor,line width= 0.4pt,line join=round,line cap=round,fill=fillColor] (226.39,196.39) circle (  2.50);

\path[draw=drawColor,line width= 0.4pt,line join=round,line cap=round,fill=fillColor] (226.39,196.39) circle (  2.50);

\path[draw=drawColor,line width= 0.4pt,line join=round,line cap=round,fill=fillColor] (179.64,206.98) circle (  2.50);

\path[draw=drawColor,line width= 0.4pt,line join=round,line cap=round,fill=fillColor] (226.39,204.04) circle (  2.50);

\path[draw=drawColor,line width= 0.4pt,line join=round,line cap=round,fill=fillColor] (226.39,204.04) circle (  2.50);

\path[draw=drawColor,line width= 0.4pt,line join=round,line cap=round,fill=fillColor] (226.39,203.45) circle (  2.50);

\path[draw=drawColor,line width= 0.4pt,line join=round,line cap=round,fill=fillColor] (179.64,211.69) circle (  2.50);

\path[draw=drawColor,line width= 0.4pt,line join=round,line cap=round,fill=fillColor] (226.39,209.93) circle (  2.50);

\path[draw=drawColor,line width= 0.4pt,line join=round,line cap=round,fill=fillColor] (226.39,209.93) circle (  2.50);

\path[draw=drawColor,line width= 0.4pt,line join=round,line cap=round,fill=fillColor] (226.39,208.16) circle (  2.50);

\path[draw=drawColor,line width= 0.6pt,dash pattern=on 4pt off 4pt ,line join=round] (142.24,196.39) -- (240.42,196.39);
\definecolor{drawColor}{gray}{0.70}

\path[draw=drawColor,line width= 0.6pt,line join=round,line cap=round] (142.24,187.73) rectangle (240.42,339.28);
\end{scope}
\begin{scope}
\path[clip] (142.24, 30.69) rectangle (240.42,182.23);
\definecolor{fillColor}{RGB}{255,255,255}

\path[fill=fillColor] (142.24, 30.69) rectangle (240.42,182.23);
\definecolor{drawColor}{gray}{0.87}

\path[draw=drawColor,line width= 0.1pt,line join=round] (142.24, 54.06) --
	(240.42, 54.06);

\path[draw=drawColor,line width= 0.1pt,line join=round] (142.24, 83.50) --
	(240.42, 83.50);

\path[draw=drawColor,line width= 0.1pt,line join=round] (142.24,112.94) --
	(240.42,112.94);

\path[draw=drawColor,line width= 0.1pt,line join=round] (142.24,142.37) --
	(240.42,142.37);

\path[draw=drawColor,line width= 0.1pt,line join=round] (142.24,171.81) --
	(240.42,171.81);

\path[draw=drawColor,line width= 0.3pt,line join=round] (142.24, 39.34) --
	(240.42, 39.34);

\path[draw=drawColor,line width= 0.3pt,line join=round] (142.24, 68.78) --
	(240.42, 68.78);

\path[draw=drawColor,line width= 0.3pt,line join=round] (142.24, 98.22) --
	(240.42, 98.22);

\path[draw=drawColor,line width= 0.3pt,line join=round] (142.24,127.65) --
	(240.42,127.65);

\path[draw=drawColor,line width= 0.3pt,line join=round] (142.24,157.09) --
	(240.42,157.09);

\path[draw=drawColor,line width= 0.3pt,line join=round] (156.26, 30.69) --
	(156.26,182.23);

\path[draw=drawColor,line width= 0.3pt,line join=round] (179.64, 30.69) --
	(179.64,182.23);

\path[draw=drawColor,line width= 0.3pt,line join=round] (203.01, 30.69) --
	(203.01,182.23);

\path[draw=drawColor,line width= 0.3pt,line join=round] (226.39, 30.69) --
	(226.39,182.23);
\definecolor{drawColor}{RGB}{0,0,0}
\definecolor{fillColor}{RGB}{0,0,0}

\path[draw=drawColor,line width= 0.4pt,line join=round,line cap=round,fill=fillColor] (156.26,175.34) circle (  2.50);

\path[draw=drawColor,line width= 0.4pt,line join=round,line cap=round,fill=fillColor] (203.01,152.38) circle (  2.50);

\path[draw=drawColor,line width= 0.4pt,line join=round,line cap=round,fill=fillColor] (203.01,174.17) circle (  2.50);

\path[draw=drawColor,line width= 0.4pt,line join=round,line cap=round,fill=fillColor] (203.01, 87.03) circle (  2.50);

\path[draw=drawColor,line width= 0.4pt,line join=round,line cap=round,fill=fillColor] (179.64, 69.37) circle (  2.50);

\path[draw=drawColor,line width= 0.4pt,line join=round,line cap=round,fill=fillColor] (226.39, 69.37) circle (  2.50);

\path[draw=drawColor,line width= 0.4pt,line join=round,line cap=round,fill=fillColor] (226.39, 69.37) circle (  2.50);

\path[draw=drawColor,line width= 0.4pt,line join=round,line cap=round,fill=fillColor] (226.39, 69.37) circle (  2.50);

\path[draw=drawColor,line width= 0.4pt,line join=round,line cap=round,fill=fillColor] (179.64, 71.72) circle (  2.50);

\path[draw=drawColor,line width= 0.4pt,line join=round,line cap=round,fill=fillColor] (226.39, 71.13) circle (  2.50);

\path[draw=drawColor,line width= 0.4pt,line join=round,line cap=round,fill=fillColor] (226.39, 70.54) circle (  2.50);

\path[draw=drawColor,line width= 0.4pt,line join=round,line cap=round,fill=fillColor] (226.39, 70.54) circle (  2.50);

\path[draw=drawColor,line width= 0.4pt,line join=round,line cap=round,fill=fillColor] (179.64, 73.49) circle (  2.50);

\path[draw=drawColor,line width= 0.4pt,line join=round,line cap=round,fill=fillColor] (226.39, 72.90) circle (  2.50);

\path[draw=drawColor,line width= 0.4pt,line join=round,line cap=round,fill=fillColor] (226.39, 72.90) circle (  2.50);

\path[draw=drawColor,line width= 0.4pt,line join=round,line cap=round,fill=fillColor] (226.39, 72.31) circle (  2.50);

\path[draw=drawColor,line width= 0.6pt,dash pattern=on 4pt off 4pt ,line join=round] (142.24, 39.34) -- (240.42, 39.34);
\definecolor{drawColor}{gray}{0.70}

\path[draw=drawColor,line width= 0.6pt,line join=round,line cap=round] (142.24, 30.69) rectangle (240.42,182.23);
\end{scope}
\begin{scope}
\path[clip] (245.92,187.73) rectangle (344.10,339.28);
\definecolor{fillColor}{RGB}{255,255,255}

\path[fill=fillColor] (245.92,187.73) rectangle (344.10,339.28);
\definecolor{drawColor}{gray}{0.87}

\path[draw=drawColor,line width= 0.1pt,line join=round] (245.92,211.11) --
	(344.10,211.11);

\path[draw=drawColor,line width= 0.1pt,line join=round] (245.92,240.54) --
	(344.10,240.54);

\path[draw=drawColor,line width= 0.1pt,line join=round] (245.92,269.98) --
	(344.10,269.98);

\path[draw=drawColor,line width= 0.1pt,line join=round] (245.92,299.42) --
	(344.10,299.42);

\path[draw=drawColor,line width= 0.1pt,line join=round] (245.92,328.86) --
	(344.10,328.86);

\path[draw=drawColor,line width= 0.3pt,line join=round] (245.92,196.39) --
	(344.10,196.39);

\path[draw=drawColor,line width= 0.3pt,line join=round] (245.92,225.82) --
	(344.10,225.82);

\path[draw=drawColor,line width= 0.3pt,line join=round] (245.92,255.26) --
	(344.10,255.26);

\path[draw=drawColor,line width= 0.3pt,line join=round] (245.92,284.70) --
	(344.10,284.70);

\path[draw=drawColor,line width= 0.3pt,line join=round] (245.92,314.14) --
	(344.10,314.14);

\path[draw=drawColor,line width= 0.3pt,line join=round] (259.94,187.73) --
	(259.94,339.28);

\path[draw=drawColor,line width= 0.3pt,line join=round] (283.32,187.73) --
	(283.32,339.28);

\path[draw=drawColor,line width= 0.3pt,line join=round] (306.69,187.73) --
	(306.69,339.28);

\path[draw=drawColor,line width= 0.3pt,line join=round] (330.07,187.73) --
	(330.07,339.28);
\definecolor{drawColor}{RGB}{0,0,0}
\definecolor{fillColor}{RGB}{0,0,0}

\path[draw=drawColor,line width= 0.4pt,line join=round,line cap=round,fill=fillColor] (259.94,331.80) circle (  2.50);

\path[draw=drawColor,line width= 0.4pt,line join=round,line cap=round,fill=fillColor] (306.69,196.39) circle (  2.50);

\path[draw=drawColor,line width= 0.4pt,line join=round,line cap=round,fill=fillColor] (306.69,194.62) circle (  2.50);

\path[draw=drawColor,line width= 0.4pt,line join=round,line cap=round,fill=fillColor] (306.69,195.80) circle (  2.50);

\path[draw=drawColor,line width= 0.4pt,line join=round,line cap=round,fill=fillColor] (283.32,304.72) circle (  2.50);

\path[draw=drawColor,line width= 0.4pt,line join=round,line cap=round,fill=fillColor] (330.07,202.27) circle (  2.50);

\path[draw=drawColor,line width= 0.4pt,line join=round,line cap=round,fill=fillColor] (330.07,202.86) circle (  2.50);

\path[draw=drawColor,line width= 0.4pt,line join=round,line cap=round,fill=fillColor] (330.07,206.98) circle (  2.50);

\path[draw=drawColor,line width= 0.4pt,line join=round,line cap=round,fill=fillColor] (283.32,255.26) circle (  2.50);

\path[draw=drawColor,line width= 0.4pt,line join=round,line cap=round,fill=fillColor] (330.07,202.27) circle (  2.50);

\path[draw=drawColor,line width= 0.4pt,line join=round,line cap=round,fill=fillColor] (330.07,204.63) circle (  2.50);

\path[draw=drawColor,line width= 0.4pt,line join=round,line cap=round,fill=fillColor] (283.32,264.68) circle (  2.50);

\path[draw=drawColor,line width= 0.4pt,line join=round,line cap=round,fill=fillColor] (330.07,212.87) circle (  2.50);

\path[draw=drawColor,line width= 0.4pt,line join=round,line cap=round,fill=fillColor] (330.07,212.87) circle (  2.50);

\path[draw=drawColor,line width= 0.4pt,line join=round,line cap=round,fill=fillColor] (330.07,215.23) circle (  2.50);

\path[draw=drawColor,line width= 0.4pt,line join=round,line cap=round,fill=fillColor] (283.32,260.56) circle (  2.50);

\path[draw=drawColor,line width= 0.4pt,line join=round,line cap=round,fill=fillColor] (330.07,230.53) circle (  2.50);

\path[draw=drawColor,line width= 0.4pt,line join=round,line cap=round,fill=fillColor] (330.07,230.53) circle (  2.50);

\path[draw=drawColor,line width= 0.4pt,line join=round,line cap=round,fill=fillColor] (330.07,231.71) circle (  2.50);

\path[draw=drawColor,line width= 0.6pt,dash pattern=on 4pt off 4pt ,line join=round] (245.92,196.39) -- (344.10,196.39);
\definecolor{drawColor}{gray}{0.70}

\path[draw=drawColor,line width= 0.6pt,line join=round,line cap=round] (245.92,187.73) rectangle (344.10,339.28);
\end{scope}
\begin{scope}
\path[clip] (245.92, 30.69) rectangle (344.10,182.23);
\definecolor{fillColor}{RGB}{255,255,255}

\path[fill=fillColor] (245.92, 30.69) rectangle (344.10,182.23);
\definecolor{drawColor}{gray}{0.87}

\path[draw=drawColor,line width= 0.1pt,line join=round] (245.92, 54.06) --
	(344.10, 54.06);

\path[draw=drawColor,line width= 0.1pt,line join=round] (245.92, 83.50) --
	(344.10, 83.50);

\path[draw=drawColor,line width= 0.1pt,line join=round] (245.92,112.94) --
	(344.10,112.94);

\path[draw=drawColor,line width= 0.1pt,line join=round] (245.92,142.37) --
	(344.10,142.37);

\path[draw=drawColor,line width= 0.1pt,line join=round] (245.92,171.81) --
	(344.10,171.81);

\path[draw=drawColor,line width= 0.3pt,line join=round] (245.92, 39.34) --
	(344.10, 39.34);

\path[draw=drawColor,line width= 0.3pt,line join=round] (245.92, 68.78) --
	(344.10, 68.78);

\path[draw=drawColor,line width= 0.3pt,line join=round] (245.92, 98.22) --
	(344.10, 98.22);

\path[draw=drawColor,line width= 0.3pt,line join=round] (245.92,127.65) --
	(344.10,127.65);

\path[draw=drawColor,line width= 0.3pt,line join=round] (245.92,157.09) --
	(344.10,157.09);

\path[draw=drawColor,line width= 0.3pt,line join=round] (259.94, 30.69) --
	(259.94,182.23);

\path[draw=drawColor,line width= 0.3pt,line join=round] (283.32, 30.69) --
	(283.32,182.23);

\path[draw=drawColor,line width= 0.3pt,line join=round] (306.69, 30.69) --
	(306.69,182.23);

\path[draw=drawColor,line width= 0.3pt,line join=round] (330.07, 30.69) --
	(330.07,182.23);
\definecolor{drawColor}{RGB}{0,0,0}
\definecolor{fillColor}{RGB}{0,0,0}

\path[draw=drawColor,line width= 0.4pt,line join=round,line cap=round,fill=fillColor] (259.94,175.34) circle (  2.50);

\path[draw=drawColor,line width= 0.4pt,line join=round,line cap=round,fill=fillColor] (306.69, 46.41) circle (  2.50);

\path[draw=drawColor,line width= 0.4pt,line join=round,line cap=round,fill=fillColor] (306.69, 69.96) circle (  2.50);

\path[draw=drawColor,line width= 0.4pt,line join=round,line cap=round,fill=fillColor] (306.69, 74.67) circle (  2.50);

\path[draw=drawColor,line width= 0.4pt,line join=round,line cap=round,fill=fillColor] (283.32,153.56) circle (  2.50);

\path[draw=drawColor,line width= 0.4pt,line join=round,line cap=round,fill=fillColor] (330.07, 79.38) circle (  2.50);

\path[draw=drawColor,line width= 0.4pt,line join=round,line cap=round,fill=fillColor] (330.07, 74.67) circle (  2.50);

\path[draw=drawColor,line width= 0.4pt,line join=round,line cap=round,fill=fillColor] (330.07, 77.61) circle (  2.50);

\path[draw=drawColor,line width= 0.4pt,line join=round,line cap=round,fill=fillColor] (283.32,109.99) circle (  2.50);

\path[draw=drawColor,line width= 0.4pt,line join=round,line cap=round,fill=fillColor] (330.07, 75.25) circle (  2.50);

\path[draw=drawColor,line width= 0.4pt,line join=round,line cap=round,fill=fillColor] (330.07, 75.84) circle (  2.50);

\path[draw=drawColor,line width= 0.4pt,line join=round,line cap=round,fill=fillColor] (283.32,115.88) circle (  2.50);

\path[draw=drawColor,line width= 0.4pt,line join=round,line cap=round,fill=fillColor] (330.07, 77.61) circle (  2.50);

\path[draw=drawColor,line width= 0.4pt,line join=round,line cap=round,fill=fillColor] (330.07, 77.02) circle (  2.50);

\path[draw=drawColor,line width= 0.4pt,line join=round,line cap=round,fill=fillColor] (330.07, 78.20) circle (  2.50);

\path[draw=drawColor,line width= 0.4pt,line join=round,line cap=round,fill=fillColor] (283.32,111.76) circle (  2.50);

\path[draw=drawColor,line width= 0.4pt,line join=round,line cap=round,fill=fillColor] (330.07, 87.03) circle (  2.50);

\path[draw=drawColor,line width= 0.4pt,line join=round,line cap=round,fill=fillColor] (330.07, 87.03) circle (  2.50);

\path[draw=drawColor,line width= 0.4pt,line join=round,line cap=round,fill=fillColor] (330.07, 88.21) circle (  2.50);

\path[draw=drawColor,line width= 0.6pt,dash pattern=on 4pt off 4pt ,line join=round] (245.92, 39.34) -- (344.10, 39.34);
\definecolor{drawColor}{gray}{0.70}

\path[draw=drawColor,line width= 0.6pt,line join=round,line cap=round] (245.92, 30.69) rectangle (344.10,182.23);
\end{scope}
\begin{scope}
\path[clip] (349.60,187.73) rectangle (447.78,339.28);
\definecolor{fillColor}{RGB}{255,255,255}

\path[fill=fillColor] (349.60,187.73) rectangle (447.78,339.28);
\definecolor{drawColor}{gray}{0.87}

\path[draw=drawColor,line width= 0.1pt,line join=round] (349.60,211.11) --
	(447.78,211.11);

\path[draw=drawColor,line width= 0.1pt,line join=round] (349.60,240.54) --
	(447.78,240.54);

\path[draw=drawColor,line width= 0.1pt,line join=round] (349.60,269.98) --
	(447.78,269.98);

\path[draw=drawColor,line width= 0.1pt,line join=round] (349.60,299.42) --
	(447.78,299.42);

\path[draw=drawColor,line width= 0.1pt,line join=round] (349.60,328.86) --
	(447.78,328.86);

\path[draw=drawColor,line width= 0.3pt,line join=round] (349.60,196.39) --
	(447.78,196.39);

\path[draw=drawColor,line width= 0.3pt,line join=round] (349.60,225.82) --
	(447.78,225.82);

\path[draw=drawColor,line width= 0.3pt,line join=round] (349.60,255.26) --
	(447.78,255.26);

\path[draw=drawColor,line width= 0.3pt,line join=round] (349.60,284.70) --
	(447.78,284.70);

\path[draw=drawColor,line width= 0.3pt,line join=round] (349.60,314.14) --
	(447.78,314.14);

\path[draw=drawColor,line width= 0.3pt,line join=round] (363.62,187.73) --
	(363.62,339.28);

\path[draw=drawColor,line width= 0.3pt,line join=round] (387.00,187.73) --
	(387.00,339.28);

\path[draw=drawColor,line width= 0.3pt,line join=round] (410.37,187.73) --
	(410.37,339.28);

\path[draw=drawColor,line width= 0.3pt,line join=round] (433.75,187.73) --
	(433.75,339.28);
\definecolor{drawColor}{RGB}{0,0,0}
\definecolor{fillColor}{RGB}{0,0,0}

\path[draw=drawColor,line width= 0.4pt,line join=round,line cap=round,fill=fillColor] (363.62,331.80) circle (  2.50);

\path[draw=drawColor,line width= 0.4pt,line join=round,line cap=round,fill=fillColor] (410.37,306.48) circle (  2.50);

\path[draw=drawColor,line width= 0.4pt,line join=round,line cap=round,fill=fillColor] (410.37,317.67) circle (  2.50);

\path[draw=drawColor,line width= 0.4pt,line join=round,line cap=round,fill=fillColor] (410.37,222.29) circle (  2.50);

\path[draw=drawColor,line width= 0.4pt,line join=round,line cap=round,fill=fillColor] (387.00,304.72) circle (  2.50);

\path[draw=drawColor,line width= 0.4pt,line join=round,line cap=round,fill=fillColor] (433.75,276.46) circle (  2.50);

\path[draw=drawColor,line width= 0.4pt,line join=round,line cap=round,fill=fillColor] (433.75,278.22) circle (  2.50);

\path[draw=drawColor,line width= 0.4pt,line join=round,line cap=round,fill=fillColor] (433.75,254.67) circle (  2.50);

\path[draw=drawColor,line width= 0.4pt,line join=round,line cap=round,fill=fillColor] (387.00,255.26) circle (  2.50);

\path[draw=drawColor,line width= 0.4pt,line join=round,line cap=round,fill=fillColor] (433.75,248.20) circle (  2.50);

\path[draw=drawColor,line width= 0.4pt,line join=round,line cap=round,fill=fillColor] (433.75,247.61) circle (  2.50);

\path[draw=drawColor,line width= 0.4pt,line join=round,line cap=round,fill=fillColor] (433.75,248.79) circle (  2.50);

\path[draw=drawColor,line width= 0.4pt,line join=round,line cap=round,fill=fillColor] (387.00,264.68) circle (  2.50);

\path[draw=drawColor,line width= 0.4pt,line join=round,line cap=round,fill=fillColor] (433.75,255.85) circle (  2.50);

\path[draw=drawColor,line width= 0.4pt,line join=round,line cap=round,fill=fillColor] (433.75,255.26) circle (  2.50);

\path[draw=drawColor,line width= 0.4pt,line join=round,line cap=round,fill=fillColor] (387.00,260.56) circle (  2.50);

\path[draw=drawColor,line width= 0.4pt,line join=round,line cap=round,fill=fillColor] (433.75,257.03) circle (  2.50);

\path[draw=drawColor,line width= 0.4pt,line join=round,line cap=round,fill=fillColor] (433.75,257.03) circle (  2.50);

\path[draw=drawColor,line width= 0.6pt,dash pattern=on 4pt off 4pt ,line join=round] (349.60,196.39) -- (447.78,196.39);
\definecolor{drawColor}{gray}{0.70}

\path[draw=drawColor,line width= 0.6pt,line join=round,line cap=round] (349.60,187.73) rectangle (447.78,339.28);
\end{scope}
\begin{scope}
\path[clip] (349.60, 30.69) rectangle (447.78,182.23);
\definecolor{fillColor}{RGB}{255,255,255}

\path[fill=fillColor] (349.60, 30.69) rectangle (447.78,182.23);
\definecolor{drawColor}{gray}{0.87}

\path[draw=drawColor,line width= 0.1pt,line join=round] (349.60, 54.06) --
	(447.78, 54.06);

\path[draw=drawColor,line width= 0.1pt,line join=round] (349.60, 83.50) --
	(447.78, 83.50);

\path[draw=drawColor,line width= 0.1pt,line join=round] (349.60,112.94) --
	(447.78,112.94);

\path[draw=drawColor,line width= 0.1pt,line join=round] (349.60,142.37) --
	(447.78,142.37);

\path[draw=drawColor,line width= 0.1pt,line join=round] (349.60,171.81) --
	(447.78,171.81);

\path[draw=drawColor,line width= 0.3pt,line join=round] (349.60, 39.34) --
	(447.78, 39.34);

\path[draw=drawColor,line width= 0.3pt,line join=round] (349.60, 68.78) --
	(447.78, 68.78);

\path[draw=drawColor,line width= 0.3pt,line join=round] (349.60, 98.22) --
	(447.78, 98.22);

\path[draw=drawColor,line width= 0.3pt,line join=round] (349.60,127.65) --
	(447.78,127.65);

\path[draw=drawColor,line width= 0.3pt,line join=round] (349.60,157.09) --
	(447.78,157.09);

\path[draw=drawColor,line width= 0.3pt,line join=round] (363.62, 30.69) --
	(363.62,182.23);

\path[draw=drawColor,line width= 0.3pt,line join=round] (387.00, 30.69) --
	(387.00,182.23);

\path[draw=drawColor,line width= 0.3pt,line join=round] (410.37, 30.69) --
	(410.37,182.23);

\path[draw=drawColor,line width= 0.3pt,line join=round] (433.75, 30.69) --
	(433.75,182.23);
\definecolor{drawColor}{RGB}{0,0,0}
\definecolor{fillColor}{RGB}{0,0,0}

\path[draw=drawColor,line width= 0.4pt,line join=round,line cap=round,fill=fillColor] (363.62,175.34) circle (  2.50);

\path[draw=drawColor,line width= 0.4pt,line join=round,line cap=round,fill=fillColor] (410.37,152.38) circle (  2.50);

\path[draw=drawColor,line width= 0.4pt,line join=round,line cap=round,fill=fillColor] (410.37,174.17) circle (  2.50);

\path[draw=drawColor,line width= 0.4pt,line join=round,line cap=round,fill=fillColor] (410.37, 87.03) circle (  2.50);

\path[draw=drawColor,line width= 0.4pt,line join=round,line cap=round,fill=fillColor] (387.00,153.56) circle (  2.50);

\path[draw=drawColor,line width= 0.4pt,line join=round,line cap=round,fill=fillColor] (433.75,128.83) circle (  2.50);

\path[draw=drawColor,line width= 0.4pt,line join=round,line cap=round,fill=fillColor] (433.75,128.83) circle (  2.50);

\path[draw=drawColor,line width= 0.4pt,line join=round,line cap=round,fill=fillColor] (433.75,107.64) circle (  2.50);

\path[draw=drawColor,line width= 0.4pt,line join=round,line cap=round,fill=fillColor] (387.00,109.99) circle (  2.50);

\path[draw=drawColor,line width= 0.4pt,line join=round,line cap=round,fill=fillColor] (433.75,102.93) circle (  2.50);

\path[draw=drawColor,line width= 0.4pt,line join=round,line cap=round,fill=fillColor] (433.75,102.34) circle (  2.50);

\path[draw=drawColor,line width= 0.4pt,line join=round,line cap=round,fill=fillColor] (433.75,102.93) circle (  2.50);

\path[draw=drawColor,line width= 0.4pt,line join=round,line cap=round,fill=fillColor] (387.00,115.88) circle (  2.50);

\path[draw=drawColor,line width= 0.4pt,line join=round,line cap=round,fill=fillColor] (433.75,108.23) circle (  2.50);

\path[draw=drawColor,line width= 0.4pt,line join=round,line cap=round,fill=fillColor] (433.75,107.64) circle (  2.50);

\path[draw=drawColor,line width= 0.4pt,line join=round,line cap=round,fill=fillColor] (433.75,107.05) circle (  2.50);

\path[draw=drawColor,line width= 0.4pt,line join=round,line cap=round,fill=fillColor] (387.00,111.76) circle (  2.50);

\path[draw=drawColor,line width= 0.4pt,line join=round,line cap=round,fill=fillColor] (433.75,109.40) circle (  2.50);

\path[draw=drawColor,line width= 0.4pt,line join=round,line cap=round,fill=fillColor] (433.75,108.81) circle (  2.50);

\path[draw=drawColor,line width= 0.6pt,dash pattern=on 4pt off 4pt ,line join=round] (349.60, 39.34) -- (447.78, 39.34);
\definecolor{drawColor}{gray}{0.70}

\path[draw=drawColor,line width= 0.6pt,line join=round,line cap=round] (349.60, 30.69) rectangle (447.78,182.23);
\end{scope}
\begin{scope}
\path[clip] ( 38.56,339.28) rectangle (136.74,370);
\definecolor{drawColor}{RGB}{0,0,0}
\definecolor{fillColor}{RGB}{255,255,255}

\path[draw=drawColor,line width= 0.6pt,line join=round,line cap=round,fill=fillColor] ( 38.56,339.28) rectangle (136.74,370);

\node[text=drawColor,anchor=base,inner sep=0pt, outer sep=0pt, scale=  0.88] at ( 87.65,358) {Scenario 1};

\node[text=drawColor,anchor=base,inner sep=0pt, outer sep=0pt, scale=  0.6] at ( 87.65,345) {$\widehat{m}_r ( \ub_k )$ \checkmark ~~~ $\widehat{p}_k $ \checkmark};

\end{scope}
\begin{scope}
\path[clip] (142.24,339.28) rectangle (240.42,370);
\definecolor{drawColor}{RGB}{0,0,0}
\definecolor{fillColor}{RGB}{255,255,255}

\path[draw=drawColor,line width= 0.6pt,line join=round,line cap=round,fill=fillColor] (142.24,339.28) rectangle (240.42,370);

\node[text=drawColor,anchor=base,inner sep=0pt, outer sep=0pt, scale=  0.88] at (191.33,358) {Scenario 2};

\node[text=drawColor,anchor=base,inner sep=0pt, outer sep=0pt, scale=  0.6] at ( 191.33,345) {$\widehat{m}_r ( \ub_k )$ \checkmark ~~~ $\widehat{p}_k $ \(\times\)};

\end{scope}
\begin{scope}
\path[clip] (245.92,339.28) rectangle (344.10,370);
\definecolor{drawColor}{RGB}{0,0,0}
\definecolor{fillColor}{RGB}{255,255,255}

\path[draw=drawColor,line width= 0.6pt,line join=round,line cap=round,fill=fillColor] (245.92,339.28) rectangle (344.10,370);

\node[text=drawColor,anchor=base,inner sep=0pt, outer sep=0pt, scale=  0.88] at (295.01,358) {Scenario 3};

\node[text=drawColor,anchor=base,inner sep=0pt, outer sep=0pt, scale=  0.6] at ( 295.01,345) {$\widehat{m}_r ( \ub_k )$ \(\times\) ~~~ $\widehat{p}_k $ \checkmark};

\end{scope}
\begin{scope}
\path[clip] (349.60,339.28) rectangle (447.78,370);
\definecolor{drawColor}{RGB}{0,0,0}
\definecolor{fillColor}{RGB}{255,255,255}

\path[draw=drawColor,line width= 0.6pt,line join=round,line cap=round,fill=fillColor] (349.60,339.28) rectangle (447.78,370);

\node[text=drawColor,anchor=base,inner sep=0pt, outer sep=0pt, scale=  0.88] at (398.69,358) {Scenario 4};

\node[text=drawColor,anchor=base,inner sep=0pt, outer sep=0pt, scale=  0.6] at (398.69,345) {$\widehat{m}_r ( \ub_k )$ \(\times\) ~~~ $\widehat{p}_k $ \(\times\)};

\end{scope}
\begin{scope}
\path[clip] (447.78,187.73) rectangle (464.35,339.28);
\definecolor{drawColor}{RGB}{0,0,0}
\definecolor{fillColor}{RGB}{255,255,255}

\path[draw=drawColor,line width= 0.6pt,line join=round,line cap=round,fill=fillColor] (447.78,187.73) rectangle (464.35,339.28);

\node[text=drawColor,rotate=-90.00,anchor=base,inner sep=0pt, outer sep=0pt, scale=  0.88] at (453.03,263.51) {RB};
\end{scope}
\begin{scope}
\path[clip] (447.78, 30.69) rectangle (464.35,182.23);
\definecolor{drawColor}{RGB}{0,0,0}
\definecolor{fillColor}{RGB}{255,255,255}

\path[draw=drawColor,line width= 0.6pt,line join=round,line cap=round,fill=fillColor] (447.78, 30.69) rectangle (464.35,182.23);

\node[text=drawColor,rotate=-90.00,anchor=base,inner sep=0pt, outer sep=0pt, scale=  0.88] at (453.03,106.46) {RSd};
\end{scope}
\begin{scope}
\path[clip] (  0.00,  0.00) rectangle (469.85,361.35);
\definecolor{drawColor}{gray}{0.70}

\path[draw=drawColor,line width= 0.3pt,line join=round] ( 52.58, 27.94) --
	( 52.58, 30.69);

\path[draw=drawColor,line width= 0.3pt,line join=round] ( 75.96, 27.94) --
	( 75.96, 30.69);

\path[draw=drawColor,line width= 0.3pt,line join=round] ( 99.33, 27.94) --
	( 99.33, 30.69);

\path[draw=drawColor,line width= 0.3pt,line join=round] (122.71, 27.94) --
	(122.71, 30.69);
\end{scope}
\begin{scope}
\path[clip] (  0.00,  0.00) rectangle (469.85,361.35);
\definecolor{drawColor}{gray}{0.30}

\node[text=drawColor,anchor=base,inner sep=0pt, outer sep=0pt, scale=  1, rotate = 45] at ( 52.58, 15.68) {$\widehat{t}_{naive}$};

\node[text=drawColor,anchor=base,inner sep=0pt, outer sep=0pt, scale=  1, rotate = 45] at ( 75.96, 15.68) {$\widehat{t}_{imp}$};

\node[text=drawColor,anchor=base,inner sep=0pt, outer sep=0pt, scale=  1, rotate = 45] at ( 99.33, 15.68) {$\widehat{t}_{NWA}$};

\node[text=drawColor,anchor=base,inner sep=0pt, outer sep=0pt, scale=  1, rotate = 45] at (122.71, 15.68) {$\widehat{t}_{qma,\pi}$};
\end{scope}
\begin{scope}
\path[clip] (  0.00,  0.00) rectangle (469.85,361.35);
\definecolor{drawColor}{gray}{0.70}

\path[draw=drawColor,line width= 0.3pt,line join=round] (156.26, 27.94) --
	(156.26, 30.69);

\path[draw=drawColor,line width= 0.3pt,line join=round] (179.64, 27.94) --
	(179.64, 30.69);

\path[draw=drawColor,line width= 0.3pt,line join=round] (203.01, 27.94) --
	(203.01, 30.69);

\path[draw=drawColor,line width= 0.3pt,line join=round] (226.39, 27.94) --
	(226.39, 30.69);
\end{scope}
\begin{scope}
\path[clip] (  0.00,  0.00) rectangle (469.85,361.35);
\definecolor{drawColor}{gray}{0.30}

\node[text=drawColor,anchor=base,inner sep=0pt, outer sep=0pt, scale=  1, rotate = 45] at (156.26, 15.68) {$\widehat{t}_{naive}$};

\node[text=drawColor,anchor=base,inner sep=0pt, outer sep=0pt, scale=  1, rotate = 45] at (179.64, 15.68) {$\widehat{t}_{imp}$};

\node[text=drawColor,anchor=base,inner sep=0pt, outer sep=0pt, scale=  1, rotate = 45] at (203.01, 15.68) {$\widehat{t}_{NWA}$};

\node[text=drawColor,anchor=base,inner sep=0pt, outer sep=0pt, scale=  1, rotate = 45] at (226.39, 15.68) {$\widehat{t}_{qma,\pi}$};
\end{scope}
\begin{scope}
\path[clip] (  0.00,  0.00) rectangle (469.85,361.35);
\definecolor{drawColor}{gray}{0.70}

\path[draw=drawColor,line width= 0.3pt,line join=round] (259.94, 27.94) --
	(259.94, 30.69);

\path[draw=drawColor,line width= 0.3pt,line join=round] (283.32, 27.94) --
	(283.32, 30.69);

\path[draw=drawColor,line width= 0.3pt,line join=round] (306.69, 27.94) --
	(306.69, 30.69);

\path[draw=drawColor,line width= 0.3pt,line join=round] (330.07, 27.94) --
	(330.07, 30.69);
\end{scope}
\begin{scope}
\path[clip] (  0.00,  0.00) rectangle (469.85,361.35);
\definecolor{drawColor}{gray}{0.30}

\node[text=drawColor,anchor=base,inner sep=0pt, outer sep=0pt, rotate = 45, scale = 1] at (259.94, 15.68) {$\widehat{t}_{naive}$};

\node[text=drawColor,anchor=base,inner sep=0pt, outer sep=0pt, rotate = 45, scale = 1] at (283.32, 15.68) {$\widehat{t}_{imp}$};

\node[text=drawColor,anchor=base,inner sep=0pt, outer sep=0pt, rotate = 45, scale = 1] at (306.69, 15.68) {$\widehat{t}_{NWA}$};

\node[text=drawColor,anchor=base,inner sep=0pt, outer sep=0pt, rotate = 45, scale = 1] at (330.07, 15.68) {$\widehat{t}_{qma,\pi}$};
\end{scope}
\begin{scope}
\path[clip] (  0.00,  0.00) rectangle (469.85,361.35);
\definecolor{drawColor}{gray}{0.70}

\path[draw=drawColor,line width= 0.3pt,line join=round] (363.62, 27.94) --
	(363.62, 30.69);

\path[draw=drawColor,line width= 0.3pt,line join=round] (387.00, 27.94) --
	(387.00, 30.69);

\path[draw=drawColor,line width= 0.3pt,line join=round] (410.37, 27.94) --
	(410.37, 30.69);

\path[draw=drawColor,line width= 0.3pt,line join=round] (433.75, 27.94) --
	(433.75, 30.69);
\end{scope}
\begin{scope}
\path[clip] (  0.00,  0.00) rectangle (469.85,361.35);
\definecolor{drawColor}{gray}{0.30}

\node[text=drawColor,anchor=base,inner sep=0pt, outer sep=0pt, scale=  1, rotate = 45] at (363.62, 15.68) {$\widehat{t}_{naive}$};

\node[text=drawColor,anchor=base,inner sep=0pt, outer sep=0pt, scale=  1, rotate = 45] at (387.00, 15.68) {$\widehat{t}_{imp}$};

\node[text=drawColor,anchor=base,inner sep=0pt, outer sep=0pt, scale=  1, rotate = 45] at (410.37, 15.68) {$\widehat{t}_{NWA}$};

\node[text=drawColor,anchor=base,inner sep=0pt, outer sep=0pt, scale=  1, rotate = 45] at (433.75, 15.68) {$\widehat{t}_{qma,\pi}$};
\end{scope}
\begin{scope}
\path[clip] (  0.00,  0.00) rectangle (469.85,361.35);
\definecolor{drawColor}{gray}{0.30}

\node[text=drawColor,anchor=base east,inner sep=0pt, outer sep=0pt, scale=  0.88] at ( 33.61,193.36) {0.00};

\node[text=drawColor,anchor=base east,inner sep=0pt, outer sep=0pt, scale=  0.88] at ( 33.61,222.79) {0.05};

\node[text=drawColor,anchor=base east,inner sep=0pt, outer sep=0pt, scale=  0.88] at ( 33.61,252.23) {0.10};

\node[text=drawColor,anchor=base east,inner sep=0pt, outer sep=0pt, scale=  0.88] at ( 33.61,281.67) {0.15};

\node[text=drawColor,anchor=base east,inner sep=0pt, outer sep=0pt, scale=  0.88] at ( 33.61,311.11) {0.20};
\end{scope}
\begin{scope}
\path[clip] (  0.00,  0.00) rectangle (469.85,361.35);
\definecolor{drawColor}{gray}{0.70}

\path[draw=drawColor,line width= 0.3pt,line join=round] ( 35.81,196.39) --
	( 38.56,196.39);

\path[draw=drawColor,line width= 0.3pt,line join=round] ( 35.81,225.82) --
	( 38.56,225.82);

\path[draw=drawColor,line width= 0.3pt,line join=round] ( 35.81,255.26) --
	( 38.56,255.26);

\path[draw=drawColor,line width= 0.3pt,line join=round] ( 35.81,284.70) --
	( 38.56,284.70);

\path[draw=drawColor,line width= 0.3pt,line join=round] ( 35.81,314.14) --
	( 38.56,314.14);
\end{scope}
\begin{scope}
\path[clip] (  0.00,  0.00) rectangle (469.85,361.35);
\definecolor{drawColor}{gray}{0.30}

\node[text=drawColor,anchor=base east,inner sep=0pt, outer sep=0pt, scale=  0.88] at ( 33.61, 36.31) {0.00};

\node[text=drawColor,anchor=base east,inner sep=0pt, outer sep=0pt, scale=  0.88] at ( 33.61, 65.75) {0.05};

\node[text=drawColor,anchor=base east,inner sep=0pt, outer sep=0pt, scale=  0.88] at ( 33.61, 95.19) {0.10};

\node[text=drawColor,anchor=base east,inner sep=0pt, outer sep=0pt, scale=  0.88] at ( 33.61,124.62) {0.15};

\node[text=drawColor,anchor=base east,inner sep=0pt, outer sep=0pt, scale=  0.88] at ( 33.61,154.06) {0.20};
\end{scope}
\begin{scope}
\path[clip] (  0.00,  0.00) rectangle (469.85,361.35);
\definecolor{drawColor}{gray}{0.70}

\path[draw=drawColor,line width= 0.3pt,line join=round] ( 35.81, 39.34) --
	( 38.56, 39.34);

\path[draw=drawColor,line width= 0.3pt,line join=round] ( 35.81, 68.78) --
	( 38.56, 68.78);

\path[draw=drawColor,line width= 0.3pt,line join=round] ( 35.81, 98.22) --
	( 38.56, 98.22);

\path[draw=drawColor,line width= 0.3pt,line join=round] ( 35.81,127.65) --
	( 38.56,127.65);

\path[draw=drawColor,line width= 0.3pt,line join=round] ( 35.81,157.09) --
	( 38.56,157.09);
\end{scope}
\begin{scope}
\path[clip] (  0.00,  0.00) rectangle (469.85,361.35);
\definecolor{drawColor}{RGB}{0,0,0}

\end{scope}
\begin{scope}
\path[clip] (  0.00,  0.00) rectangle (469.85,361.35);
\definecolor{drawColor}{RGB}{0,0,0}

\end{scope}
\end{tikzpicture}

}
	\caption[]{
		Bias and standard deviation of total estimators relative to the true total 
        for the simulated data under scenarios~1 to 4 with four prediction methods 
        for the working model and three estimation methods for the response model. 
        For reference, the estimator $\widehat{t}_{\pi}$ (unavailable under nonresponse) 
        has a RB of~0 and a RSd of~0.05.
		\label{plotres}}
	\label{fig:plot1}
\end{figure}

In scenario 1, both the response and working models fit well the data. 
In this case, all estimators perform well, except for the naive estimator, 
with RB values close to zero and RSd values of about 5\%.
 
In scenario~2, the response model is misspecified. Estimators $\widehat{t}_{imp}$ and $\widehat{t}_{qma,\pi}$ perform the best among all considered estimators. Even if the quasi-model-assisted estimator $\widehat{t}_{qma,\pi}$ uses the misspecified response model, it performs as well as the imputed estimator $\widehat{t}_{imp}$ which is unaffected by the misspecification as it does not rely on the estimated response model. Both estimators show comparable relative bias (RB less than $\approx 2.5\%$) and relative standard deviation (RSd less than $\approx 6\%$).

In scenario 2, the figures in Table~\ref{table1:simulateddata} reveal that the NWA estimator $\widehat{t}_{NWA}$ performs better when the response probabilities are estimated using the $K$-nearest neighbors method (RB of 4.4\%, RSd of 8.1\%) than when the response probabilities are estimated using calibration or maximum likelihood (RB and RSd of about 20\%). This difference is probably explained by the fact that $K$-nearest neighbor does not rely on an assumed functional form for the response probabilities and instead estimates it non-parametrically. Calibration or maximum likelihood rely on an assumed functional form for the response probabilities that is misspecified in this scenario. With the $K$-nearest neighbor, only the variables are misspecified, with calibration and maximum likelihood the functional form and the auxiliary variables are. We noticed a similar phenomenon in terms of MAE, see Table~\ref{table2}.

In scenario~3, the working model is misspecified. Estimator $\widehat{t}_{NWA}$ provides the best results, which is explained by the fact that the response model is well specified. The quasi-model-assisted estimator $\widehat{t}_{qma,\pi}$ shows overall good performance, depending on the method used to estimate the working model, with RB between 1\% and 6\% and RSd between 6\% and 8\%. In contrast, the imputed estimator $\widehat{t}_{imp}$ exhibits larger RB (10\%--18\%) and RSd (12\%--19\%) due to the misspecified working model.

Estimator $\widehat{t}_{qma ,\pi}$ shows a RB overall smaller in scenario~2 than in scenario~3, with a RB between 0\% and 2\% in the former and between 1\% and 6\% in the latter. This difference may be explained by the fact that the conditions under which $\widehat{t}_{qma, \pi}$ is asymptotically unbiased are stronger when the working model is misspecified (scenario 3) than when the response model is misspecified (scenario 4). More details are given in Appendix~\ref{Appendix:DR}.

Finally, in scenario~4, both the response and working models are misspecified. In this case, the quasi-model-assisted estimator $\widehat{t}_{qma,\pi}$ generally performs comparably to $\widehat{t}_{imp}$ and outperforms $\widehat{t}_{NWA}$ in terms of RB and RSd. %, as the latter two rely on only one of the two models. 
Interestingly, in this scenario $\widehat{t}_{NWA}$ 
performs the best when the response probabilities are estimated using the $K$-nearest neighbors method. A possible reason is that this method is based on a non-parametric estimation of the functional function for the response probabilities and relies on only one wrong model (whereas $\widehat{t}_{qma,\pi}$ relies on two wrong models).

The general conclusion of this simulation study is that the quasi-model-assisted estimator $\widehat{t}_{qma,\pi}$ performs as well as or better than $\widehat{t}_{NWA}$, $\widehat{t}_{imp}$, and $\widehat{t}_{naive}$, regardless of whether one or both of the response and working models are misspecified. This class of estimators thus provides robustness against model misspecification and greater reliability for estimating population total.

%Estimator $\widehat{t}_{m_r,\widehat{p}}$ works as well or better even when the working model or the response model is not well specified.
%This implies that the use of the model-assisted estimator $\widehat{t}_{m_r,\widehat{p}}$ performs better than an estimator based only on nonresponse probabilities or on a working model, even if one underlying model, i.e. nonresponse or predicting models, is not appropriate.
%Finally, if the two postulated models are misspecified, the model-assisted estimator $\widehat{t}_{m_r,\widehat{p}}$ works at least as well as the others.

% ----------------------------------------------------------------
\section{Discussion}\label{section:discussion}
% ----------------------------------------------------------------

% We propose and study a new class of estimators. The novelty is that we adapt model-assisted total estimators to missing at random data building on the idea of nonresponse weighting adjustment. The new class of estimators that we propose is vast and includes many different estimators. Some well-known estimators belong to this class. Each estimator in this class is based on a working model and a response model. A vast choice of statistical learning methods is available to estimate either of these models. We give some examples.
% For some of these choices, the resulting estimator can be written as a weighted estimator. We show cases in which the resulting weights are calibrated to the total of the auxiliary variables. We conduct a simulation study to empirically study the performance of our estimator. The results of this study confirm that our estimator generally outperforms competing estimators, even when the one or both of the underlying models is or are misspecified. Further work includes the study of our estimator under other working models as well as the extension to non-missing at random data.

We present and study a class of estimators, adapting model-assisted total estimators to handle missing at random data via nonresponse weighting adjustments. This class is broad, encompassing many estimators, including well-known ones. Each estimator relies on a working model and a response model, which can be estimated using a variety of statistical learning methods. For certain choices, the estimator reduces to a weighted estimator with weights calibrated to the total of the auxiliary variables. A simulation study evaluates the performance of the proposed estimator, showing that it generally outperforms competitors, even when one or both underlying models are misspecified. Future work includes exploring its performance under alternative working models and extending it to missing not at random data.

% ----------------------------------------------------------------
\section*{Acknowledgements}
% ----------------------------------------------------------------

This work was partially funded by the Swiss Federal Statistical Office. 
%The adress of the author Esther Eustache during the major part of this research was “Institute of Statistics, University of Neuchâtel, Av. de Bellevaux 51, 2000 Neuchâtel, Switzerland”.
The views expressed in this paper are those of the authors solely.

\bibliographystyle{apalike}
\bibliography{bib2}

\begin{thebibliography}{}

\bibitem[Beaumont, 2005]{bea:05a}
Beaumont, J.-F. (2005).
\newblock Calibrated imputation in surveys under a quasi-model-assisted approach.
\newblock {\em Journal of the Royal Statistical Society. Series B}, 67:445--458.

\bibitem[Breidt et~al., 2005]{bre:cla:ops:05}
Breidt, F.~J., Claeskens, G., and Opsomer, J.~D. (2005).
\newblock Model-assisted estimation for complex surveys using penalized splines.
\newblock {\em Biometrika}, 92:831--846.

\bibitem[Breidt and Opsomer, 2017]{bre:ops:17:modelassist}
Breidt, F.~J. and Opsomer, J. (2017).
\newblock Model-assisted survey estimation with modern prediction techniques.
\newblock {\em Statistical Science}, 32:190--205.

\bibitem[Breidt et~al., 2007]{bre:ops:joh:ran:07}
Breidt, F.~J., Opsomer, J., Johnson, A., and Ranalli, M. (2007).
\newblock Semiparametric model-assisted estimation for natural resource surveys.
\newblock {\em Survey methodology}, 33:35--44.

\bibitem[Breidt and Opsomer, 2000]{bre:ops:00}
Breidt, F.~J. and Opsomer, J.~D. (2000).
\newblock Local polynomial regression estimators in survey sampling.
\newblock {\em Annals of Statistics}, 28:1026--1053.

\bibitem[Brick, 2013]{bri:13}
Brick, J.~M. (2013).
\newblock Unit nonresponse and weighting adjustments: A critical review.
\newblock {\em Journal of Official Statistics}, 29(3):329--353.

\bibitem[Brick and Jones, 2008]{bri:jon:08}
Brick, J.~M. and Jones, M.~E. (2008).
\newblock Propensity to respond and nonresponse bias.
\newblock {\em Metron}, 66(1):51--73.

\bibitem[Da~Silva and Opsomer, 2006]{das:ops:06}
Da~Silva, D.~N. and Opsomer, J.~D. (2006).
\newblock A kernel smoothing method of adjusting for unit non-response in sample surveys.
\newblock {\em The Canadian Journal of Statistics}, 34(4):563--579.

\bibitem[Da~Silva and Opsomer, 2009]{das:ops:09}
Da~Silva, D.~N. and Opsomer, J.~D. (2009).
\newblock Nonparametric propensity weighting for survey nonresponse through local polynomial regression.
\newblock {\em Survey Methodology}, 35(2):165--176.

\bibitem[Dagdoug et~al., 2022]{dag:gog:haz:22:model_assisted}
Dagdoug, M., Goga, C., and Haziza, D. (2022).
\newblock Model-assisted estimation through random forests in finite population sampling.
\newblock {\em To appear in the Journal of the American Statistical Association}, pages 1--18.

\bibitem[Deville, 2002]{dev:02}
Deville, J.-C. (2002).
\newblock La correction de la nonr\'eponse par calage g\'en\'eralis\'e.
\newblock In {\em Actes des Journ\'ees de M\'ethodologie Statistique}, Paris. Insee-M\'ethodes.

\bibitem[Deville and Dupont, 1993]{dev:dup:93}
Deville, J.-C. and Dupont, F. (1993).
\newblock Non-r\'eponse: principes et m\'ethodes.
\newblock In {\em Actes des Journ\'ees de M\'ethodologie Statistique}, pages 53--70, INSEE, Paris.

\bibitem[Deville and S\"arndal, 1992]{dev:sar:92}
Deville, J.-C. and S\"arndal, C.-E. (1992).
\newblock Calibration estimators in survey sampling.
\newblock {\em Journal of the American Statistical Association}, 87:376--382.

\bibitem[Deville et~al., 1993]{dev:sar:sau:93}
Deville, J.-C., S\"arndal, C.-E., and Sautory, O. (1993).
\newblock Generalized raking procedure in survey sampling.
\newblock {\em Journal of the American Statistical Association}, 88:1013--1020.

\bibitem[Dupont, 1993]{dup:93}
Dupont, F. (1993).
\newblock Calage et redressement de la non-r\'eponse totale : validit\' e de la pratique courante de redressement et comparaison des m\'ethodes alternatives pour l'enqu\^ete sur la consommation alimentaire de 1989.
\newblock In {\em Actes des Journ\'ees de M\'ethodologie Statistique}, pages 9--42, INSEE, Paris.

\bibitem[Ekholm and Laaksonen, 1991]{ekh:laa:91}
Ekholm, A. and Laaksonen, S. (1991).
\newblock Weighting via response modeling in the finish household budget survey.
\newblock {\em Journal of Official Statistics}, 3:325--337.

\bibitem[Firth and Bennett, 1998]{fir:ben:98}
Firth, D. and Bennett, K.~E. (1998).
\newblock Robust models in probability sampling.
\newblock {\em Journal of the Royal Statistical Society. Series B (Statistical Methodology)}, 60(1):3--21.

\bibitem[Folsom, 1991]{fol:91}
Folsom, R.~E. (1991).
\newblock Exponential and logistic weight adjustments for sampling and nonresponse error reduction.
\newblock In {\em ASA Proceedings Of The Section On Social Statistics}, pages 197--202.

\bibitem[Fuller, 2009]{ful:09:book:sampling}
Fuller, W.~A. (2009).
\newblock {\em Sampling statistics}.
\newblock John Wiley \& Sons.

\bibitem[Fuller et~al., 1994]{ful:lou:bak:94}
Fuller, W.~A., Loughin, M.~M., and Baker, H.~D. (1994).
\newblock Regression weighting in the presence of nonresponse with application to the 1987--1988 nationwide food consumption survey.
\newblock {\em Survey Methodology}, 20:75--85.

\bibitem[Hansen and Hurwitz, 1946]{han:hur:46}
Hansen, M.~H. and Hurwitz, W.~N. (1946).
\newblock The problem of non-response in sample surveys.
\newblock {\em Journal of the American Statistical Association}, 41:517--529.

\bibitem[Hasler, 2023]{has:23}
Hasler, C. (2023).
\newblock Inference from sampling with response probabilities estimated via calibration.
\newblock Technical report, University of Neuch{\^a}tel.

\bibitem[Haziza and Beaumont, 2017]{haz:bea:17:weights:review}
Haziza, D. and Beaumont, J.-F. (2017).
\newblock {Construction of Weights in Surveys: A Review}.
\newblock {\em Statistical Science}, 32(2):206 -- 226.

\bibitem[Horvitz and Thompson, 1952]{hor:tho:52}
Horvitz, D.~G. and Thompson, D.~J. (1952).
\newblock A generalization of sampling without replacement from a finite universe.
\newblock {\em Journal of the American Statistical Association}, 47:663--685.

\bibitem[Iannacchione et~al., 1991]{ian:mil:fol:1991}
Iannacchione, V.~G., Milne, J.~G., and Folsom, R.~E. (1991).
\newblock Response probability weight adjustments using logistic regression.
\newblock In {\em In Proceedings of the Survey Research Methods Section, American Statistical Association}, pages 637--642.

\bibitem[Kang and Schafer, 2007]{kan:sch:07}
Kang, J.~D.~Y. and Schafer, J.~L. (2007).
\newblock Demystifying double robustness: a comparison of alternative strategies for estimating a population mean from incomplete data.
\newblock {\em Statistical Science}, 22:523--539.

\bibitem[Kim and Haziza, 2014]{kim:haz:14}
Kim, J.~K. and Haziza, D. (2014).
\newblock Doubly robust inference with missing data in survey sampling.
\newblock {\em Statistica Sinica}, 24(1):375--394.

\bibitem[Kim and Kim, 2007]{kim:kim:07}
Kim, J.~K. and Kim, J. (2007).
\newblock Nonresponse weighting adjustment using estimated response probability.
\newblock {\em The Canadian Journal of Statistics / La Revue Canadienne de Statistique}, 35(4):501--514.

\bibitem[Kim and Riddles, 2012]{kim:rid:12}
Kim, J.~K. and Riddles, M.~K. (2012).
\newblock Some theory for propensity-score-adjustment estimators in survey sampling.
\newblock {\em Survey Methodology}, 38(2):157--165.

\bibitem[Kott, 2006]{kot:06}
Kott, P.~S. (2006).
\newblock Using calibration weighting to adjust for nonresponse and coverage errors.
\newblock {\em Survey Methodology}, 32(2):133--142.

\bibitem[Kott and Chang, 2010]{kot:cha:10}
Kott, P.~S. and Chang, T. (2010).
\newblock Using calibration weighting to adjust for nonignorable unit nonresponse.
\newblock {\em Journal of the American Statistical Association}, 105(491):1265--1275.

\bibitem[Kott and Liao, 2012]{kot:lia:12}
Kott, P.~S. and Liao, D. (2012).
\newblock Providing double protection for unit nonresponse with a nonlinear calibration-weighting routine.
\newblock {\em Survey Research Methods}, 6(2):105--111.

\bibitem[Lee et~al., 2002]{lee:ran:sar:02}
Lee, H., Rancourt, E., and S\"{a}rndal, C.-E. (2002).
\newblock Variance estimation from survey data under single imputation.
\newblock In Groves, R.~M., Dillman, D.~A., Eltinge, J.~L., and Little, R. J.~A., editors, {\em In Survey Nonresponse}, pages 315--328, New York. Wiley.

\bibitem[Lesage et~al., 2019]{les:haz:dha:19}
Lesage, E., Haziza, D., and D'Haultfoeuille, X. (2019).
\newblock A cautionary tale on instrumental calibration for the treatment of nonignorable unit nonresponse in surveys.
\newblock {\em Journal of the American Statistical Association}, 114(526):906--915.

\bibitem[Lundstr\"om and S\"arndal, 1999]{lun:sar:99}
Lundstr\"om, S. and S\"arndal, C.-E. (1999).
\newblock Calibration as a standard method for treatment of nonresponse.
\newblock {\em Journal of Official Statistics}, 15:305--327.

\bibitem[Niyonsenga, 1994]{niy:94}
Niyonsenga, T. (1994).
\newblock Nonparametric estimation of response probabilities in sampling theory.
\newblock {\em Survey Methodology}, 20(2):177--184.

\bibitem[Niyonsenga, 1997]{niy:97}
Niyonsenga, T. (1997).
\newblock Response probability estimation.
\newblock {\em Journal of Statistical Planning and Inference}, 59:111--126.

\bibitem[Opsomer et~al., 2007]{ops:bre:moi:kau:07}
Opsomer, J., Breidt, F.~J., Moisen, G., and Kauermann, G. (2007).
\newblock Model-assisted estimation of forest resources with generalized additive models.
\newblock {\em Journal of the American Statistical Association}, 102:400--409.

\bibitem[Robins et~al., 1994]{rob:rot:zha:94}
Robins, J.~M., Rotnitzky, A., and Zhao, L.~P. (1994).
\newblock Estimation of regression coefficients when some regressors are not always observed.
\newblock {\em Journal of the American Statistical Association}, 89(427):846--866.

\bibitem[Robinson and S\"arndal, 1983]{rob:sar:83}
Robinson, P.~M. and S\"arndal, C.-E. (1983).
\newblock Asymptotic properties of the generalized regression estimator in probability sampling.
\newblock {\em Sankhy$\bar{a}$}, B45:240--248.

\bibitem[S\"arndal, 1980]{sar:80}
S\"arndal, C.-E. (1980).
\newblock On $\pi$-inverse weighting versus best linear unbiased weighting in probability sampling.
\newblock {\em Biometrika}, 67:639--650.

\bibitem[S\"arndal and Lundstr\"om, 2005]{sar:lun:05}
S\"arndal, C.-E. and Lundstr\"om, S. (2005).
\newblock {\em Estimation in surveys with nonresponse}.
\newblock Wiley, New York.

\bibitem[S\"arndal and Swensson, 1987]{sar:swe:87}
S\"arndal, C.-E. and Swensson, B. (1987).
\newblock A general view of estimation for two phases of selection with applications to two-phase sampling and nonresponse.
\newblock {\em International Statistical Review}, 55(3):279--294.

\bibitem[S\"arndal et~al., 1992]{sar:swe:wre:92}
S\"arndal, C.-E., Swensson, B., and Wretman, J.~H. (1992).
\newblock {\em Model Assisted Survey Sampling}.
\newblock Springer, New York.

\bibitem[S\"arndal and Wright, 1984]{sar:wri:84}
S\"arndal, C.-E. and Wright, R.~L. (1984).
\newblock Cosmetic form of estimators in survey sampling.
\newblock {\em Scandinavian Journal of Statistics}, 11:146--156.

\end{thebibliography}
% ----------------------------------------------------------------
\begin{comment}

\end{comment}

% ----------------------------------------------------------------
\section*{Appendices}
\appendix
\label{section:appendices}

\section{Double Robustness}\label{Appendix:DR}

Consider $1/p^*_k$ and $m^*(\xb_k)$ the $pq$-probability limits of $1/\widehat{p}_k$ and $\widehat{m}_r(\xb_k)$, respectively, where the reference probability is the one induced jointly by the sampling and response processes.
We define the hypothetical estimator
\begin{equation}
	t^*_{qma} = \sum_{k \in U} m^*(\xb_k) + \sum_{k \in s_r} \dfrac{y_k - m^*(\xb_k)}{\pi_k p^*_k}, \label{estimator:diff:star}
\end{equation}
which cannot be generally computed but is useful for characterizing the asymptotic properties of $\widehat{t}_{qma}$.
We can write the difference
\begin{align}
	\widehat{t}_{qma} - t^*_{qma} &= \sum_{k \in U} \left\{\widehat{m}_r(\xb_k)  -m^*(\xb_k)\right\} + \sum_{k \in U} y_k \left(\frac{a_k r_k}{\pi_k \widehat{p}_k} - \frac{a_k r_k}{\pi_k p^*_k}\right)\\
    &-\left\{\sum_{k \in U} \widehat{m}_r(\xb_k)\frac{a_k r_k}{\pi_k \widehat{p}_k}  - \sum_{k \in U}m^*(\xb_k)  \frac{a_k r_k}{\pi_k p^*_k}\right\}.
\end{align}
All three terms converge to 0 in $pq$-probability, and therefore $\widehat{t}_{qma}$ and $t^*_{qma}$ share the same $pq$-probability limit.
Moreover, 
\begin{align}
    t^*_{qma}-t &= \sum_{k \in U}\left\{y_k - m^*(\xb_k)\right\}\left(\frac{a_k r_k}{\pi_k p^*_k}-1\right).
\end{align}

If the working model is well specified, then 
%we can expect the first term to converge towards 0, at least in an average sense. Indeed, if the working model is correct, then 
$m^*(\xb_k)=m(\xb_k)$. Moreover, we have $\E_\xi(y_k|\xb_k) = m(\xb_k)$. 
Under the assumptions of ignorable nonresponse and non-informative sampling, we then have
\begin{align}
    \E_{\xi pq}(t^*_{qma}-t) &= \sum_{k \in U}\E_{\xi pq}\left[\left\{y_k - m(\xb_k)\right\}\left(\frac{a_k r_k}{\pi_k p^*_k}-1\right)\right]\\
    &= \sum_{k \in U}\left\{\E_{\xi}(y_k|\xb_k) - m(\xb_k)\right\}\E_{pq}\left(\frac{a_k r_k}{\pi_k p^*_k}-1\right)\\
    &= 0.
\end{align}
Thus, $t^*_{qma}$ is $\xi pq$–unbiased. Since $\widehat{t}_{qma}$ and $t^*_{qma}$ share the same $pq$–probability limit, $\widehat{t}_{qma}$ is asymptotically $\xi pq$–unbiased even when the response model is misspecified. 

If the response model is well specified, then $p^*_k = p_k$. We obtain
\begin{align}
    \E_{pq}(t^*_{qma}-t) &= \sum_{k \in U}\E_{pq}\left[\left\{y_k -  m^*(\xb_k)\right\}\left(\frac{a_k r_k}{\pi_k p_k}-1\right)\right].
\end{align}
Since $y_k$ is non-random with respect to $pq$, this becomes
\begin{align}
    \E_{pq}(t^*_{qma}-t) &= -\sum_{k \in U} \frac{1}{\pi_k p_k}\cov_{pq}\left\{a_k r_k,m^*(\xb_k) \right\}.
\end{align}
Thus, $t^*_{qma}$ is $pq$-unbiased when $m^*(\xb_k)$ is independent from $a_k r_k$ for all $k \in U$. 
In this case, $\widehat{t}_{qma}$ is asymptotically $pq$-unbiased when the response model is well specified, even if the working model is misspecified.

These elements shows that the hypothetical estimator $t^*_{qma}$, and therefore $\widehat{t}_{qma}$ is doubly robust: it is asymptotically unbiased when either the working model is well specified, or the response model is well specified (with the additional condition that $m^*(\xb_k)$ is independent from $a_k r_k$ when the working model is misspecified).

\section{Simulation Study}
% % ----------------------------------------------------------------
\subsection{Model Construction Details}
\label{appendix:detailofthemodel}

All simulations are conducted using the logiciel R.
Four prediction methods were used in the simulations to estimate the regression function $m$: generalized linear regression, local polynomial regression, $K$-nearest neighbors, and random forest.
The generalized linear regression and the local polynomial regression were fitted using functions \texttt{lm} and \texttt{loess}, respectively, from the R package \texttt{stats}, with all parameters left at their default values.
The $K$-nearest neighbors model was constructed using Euclidean distance and $K=5$.
The random forest model was fitted using the \texttt{randomForest} function from the R package \texttt{randomForest}, with the number of trees set to $ntree = 500$ and all other parameters left at their default values.

Three methods are used to estimate the response probabilities $p_k$: calibration, maximum likelihood estimation, and the $K$-nearest neighbors method.
For calibration, the response probabilities were defined through the estimating Equation~\eqref{estimating:eqn:calib2} using a logistic model. These probabilities were then estimated by solving the equation with the \texttt{optim} function from the R package \texttt{stats}, using the \texttt{BFGS} optimization method and default settings.
Similarly, for maximum likelihood estimation, parameters were defined through Equation~\eqref{eqn:mle:logistic} with $c_k = 1$, and the optimization problem was solved using the \texttt{optim} function with the \texttt{BFGS} method and default parameters.
For $K$-nearest neighbors, the response probability of a unit $k \in s$ is estimated as the proportion of respondents among its $K$-nearest neighbors in terms of the values of the auxiliary variables.
The Euclidean distance was used to define the neighborhood, and the value of $K$ was set to~10.

\subsection{Response and Working Models Quality Across the Different Scenarios}
\label{appendix:model:quality}

The response model shows a better fit in  scenarios 1 and 3 than in scenarios 2 and 4, with a MAE between 0.03 and 0.11 in the former and between 0.16 and 0.19 in the latter.
In scenarios 1 and 3, the MAE of $K$-nearest neighbor is larger than the MAE of calibration and maximum likelihood. This is expected since with calibration not only the auxiliary variables are correctly specified, but also the functional form of the response probabilities. With $K$-nearest neighbor, no particular functional form is assumed as it is estimated non-parametrically. In scenarios 2 and 4, however, the MAE of $K$-nearest neighbor is smaller than the MAE of calibration and maximum likelihood. In this case, $K$-nearest neighbor provides the best fit because the functional form for the response probabilities is estimated non-parametrically. Calibration and maximum likelihood are based on an assumed functional form that is incorrect.

The working models show a better fit in scenarios 1 and 2 than in scenarios 3 and 4 with a MRPE between 0.10 and 0.12 in the former and between 0.26 and 0.37 in the latter.
In scenarios 3 and 4, the MRPE of the linear model is greater than the MRPE of the others. 
In this case, the model is constrained to be linear, whereas the true relationship is nonlinear. This restriction prevents the model from adapting to the data, resulting in lower efficiency compared to a model that allows a more flexible functional form, such as polynomial linear regression, $K$-nearest neighbors, and random forest.

\subsection{Detailed Results of The Simulations}
\label{qma:appendix:D}
The results of the simulations for the simulated data are presented in Table~\ref{table1:simulateddata}. 

\begin{table}[!htb]
    \begin{subtable}[h]{0.42\textwidth}
        \centering
        \resizebox{6cm}{!}{
        \begin{tabular}[t]{l|ccc|c}
            \toprule
            \diagbox[width=5em,height=3em]{$\widehat{m}_r$}{$\widehat{p}_k$} & Calib & MLE & $K$-nn & $\varnothing$\\
            \hline
             & \multicolumn{3}{c|}{$\widehat{t}_{qma,\pi}$} & $\widehat{t}_{imp}$ \\ 
             \vspace{-3mm}
             GREG & 0.000 & 0.000 & 0.000 & -0.002\\
             & (0.051) & (0.051) & (0.051) & (0.051)\\ \vspace{-3mm}
            Poly & 0.000 & 0.000 & 0.000 & -0.001\\
            & (0.051) & (0.051) & (0.051) & (0.051)\\ \vspace{-3mm}
            $K$-nn & 0.005 & 0.005 & 0.005 & 0.018\\
            & (0.052) & (0.052) & (0.052) & (0.051)\\ \vspace{-3mm}
            RF & 0.014 & 0.014 & 0.014 & 0.026\\
            & (0.053) & (0.054) & (0.054) & (0.058)\\
            \hline
            & \multicolumn{3}{c|}{$\widehat{t}_{qma,\pi}$} & $\widehat{t}_{naive}$ \\ \vspace{-3mm}
            $\varnothing$ & 0.000 & -0.003 & -0.001 & 0.230\\
            & (0.051) & (0.052) & (0.060) & (0.231)\\
            \bottomrule
        \end{tabular}}
        \vspace{-0.4cm}
        \caption{Scenario 1 ($\widehat{m}_r ( \ub_k )$ \checkmark ~ $\widehat{p}_k $ \checkmark)}
       \label{table1:a}
    \end{subtable}
    \hspace{1cm}
    \begin{subtable}[h]{0.42\textwidth}
        \centering
        \resizebox{6cm}{!}{
        \begin{tabular}[t]{l|ccc|c}
            \toprule
             \diagbox[width=5em,height=3em]{$\widehat{m}_r$}{$\widehat{p}_k$} & Calib & MLE & $K$-nn & $\varnothing$\\
            \hline
             & \multicolumn{3}{c|}{$\widehat{t}_{qma, \pi}$} & $\widehat{t}_{imp}$ \\ \vspace{-3mm}
             GREG & -0.001 & -0.001 & -0.001 & -0.002\\
             & (0.051) & (0.051) & (0.051) & (0.051)\\ \vspace{-3mm}
            Poly & 0.000 & 0.000 & -0.001 & -0.001\\
            & (0.051) & (0.051) & (0.051) & (0.051)\\ \vspace{-3mm}
            $K$-nn & 0.013 & 0.013 & 0.012 & 0.018\\
            & (0.053) & (0.053) & (0.053) & (0.055)\\ \vspace{-3mm}
            RF & 0.023 & 0.023 & 0.020 & 0.026\\
            & (0.056) & (0.057) & (0.056) & (0.058)\\
            \hline 
            & \multicolumn{3}{c|}{$\widehat{t}_{NWA}$} & $\widehat{t}_{naive}$ \\ \vspace{-3mm}
            $\varnothing$ & 0.187 & 0.206 & 0.044 & 0.230\\
            & (0.201) & (0.229) & (0.081) & (0.231)\\
            \bottomrule
        \end{tabular}}
        \vspace{-0.4cm}
       \caption{Scenario 2 ($\widehat{m}_r ( \ub_k )$ \checkmark ~ $\widehat{p}_k $ \(\times\))}
       \label{table1:a}
    \end{subtable}
    \caption{Bias and standard deviation (in brackets) of total estimators relative to the true total for scenarios 1 and 2, with four prediction methods (generalized linear regression (GREG), local polynomial regression (poly), $K$-nearest neighbors with $K=5$ ($K$-nn), and random forest (RF)) and three estimation methods for the response probabilities (calibration (Calib), maximum likelihood (MLE), and $K$-nearest neighbors with $K=10$ ($K$-nn)). \label{table1:simulateddata}}
\end{table}

\begin{table}[!htb]
    \begin{subtable}[h]{0.42\textwidth}
        \centering
        \resizebox{6.5cm}{!}{
        \begin{tabular}[t]{l|ccc|c}
            \toprule
             \diagbox[width=5em,height=3em]{$\widehat{m}_r$}{$\widehat{p}_k$} & Calib & MLE & $K$-nn & $\varnothing$\\
            \hline
             & \multicolumn{3}{c|}{$\widehat{t}_{qma, \pi}$} & $\widehat{t}_{imp}$ \\ \vspace{-3mm}
            GREG & 0.009 & 0.011 & 0.018 & 0.184\\
            & (0.055) & (0.060) & (0.065) & (0.194)\\ \vspace{-3mm}
            Poly & 0.008 & 0.010 & 0.014 & 0.100\\
            &(0.057) & (0.060) & (0.062) & (0.120)\\ \vspace{-3mm}
            $K$-nn & 0.027 & 0.028 & 0.032 & 0.116\\
            &(0.061) & (0.064) & (0.066) & (0.130)\\ \vspace{-3mm}
            RF & 0.057 & 0.058 & 0.060 & 0.109\\
            & (0.079) & (0.081) & (0.083) & (0.123)\\ 
            \hline
            & \multicolumn{3}{c|}{$\widehat{t}_{NWA}$} & $\widehat{t}_{naive}$ \\ \vspace{-3mm}
            $\varnothing$ & 0.000 & -0.003 & -0.001 & 0.230\\
            & (0.051) & (0.052) & (0.060) & (0.231) \\
            \bottomrule
        \end{tabular}}
        \vspace{-0.4cm}
       \caption{Scenario 3 ($\widehat{m}_r ( \ub_k )$ \(\times\) ~ $\widehat{p}_k $ \checkmark)}
       \label{table1:a}
    \end{subtable}
    \hspace{1cm}
    \begin{subtable}[h]{0.42\textwidth}
        \centering
        \resizebox{6.5cm}{!}{
        \begin{tabular}[t]{l|ccc|c}
            \toprule
             \diagbox[width=5em,height=3em]{$\widehat{m}_r$}{$\widehat{p}_k$} & Calib & MLE & $K$-nn & $\varnothing$\\
            \hline
             & \multicolumn{3}{c|}{$\widehat{t}_{qma, \pi}$} & $\widehat{t}_{imp}$ \\ \vspace{-3mm}
             GREG & 0.137 & 0.139 & 0.099 & 0.184\\
             & (0.150) & (0.152) & (0.116) & (0.194) \\ \vspace{-3mm}
            Poly & 0.088 & 0.087 & 0.089 & 0.100\\
            & (0.108) & (0.107) & (0.108) & (0.120)\\ \vspace{-3mm}
            $K$-nn & 0.101 & 0.100 & 0.099 & 0.116\\
            & (0.116) & (0.116) & (0.115) & (0.130)\\ \vspace{-3mm}
            RF & 0.103 & 0.103 & 0.099 & 0.109\\
            &(0.118) & (0.118) & (0.115) & (0.123)\\
            \hline
            & \multicolumn{3}{c|}{$\widehat{t}_{NWA}$} & $\widehat{t}_{naive}$ \\ \vspace{-3mm}
            $\varnothing$ & 0.187 & 0.206 & 0.044 & 0.230\\
            &(0.201) & (0.229) & (0.081) & (0.231)\\
            \bottomrule
        \end{tabular}}
        \vspace{-0.4cm}
       \caption{Scenario 4 ($\widehat{m}_r ( \ub_k )$ \(\times\) ~ $\widehat{p}_k $ \(\times\))}
       \label{table1:a}
    \end{subtable}
    \caption{Bias and standard deviation of total estimators relative to the true total for scenarios 3 and 4, with four prediction methods (generalized linear regression (GREG), local polynomial regression (poly), $K$-nearest neighbors with $K=5$ ($K$-nn), and random forest (RF)) and three estimation methods for the response probabilities (calibration (Calib), maximum likelihood (MLE), and $K$-nearest neighbors with $K=10$ ($K$-nn)). \label{table2:simulateddata}}
\end{table}

\end{document}